\documentclass[AMA,STIX1COL]{WileyNJD-v2}
\usepackage{moreverb}
\usepackage{csquotes}
\usepackage{graphicx,color}
\usepackage{float}
\usepackage[caption=false]{subfig}
\usepackage{amsmath}
\usepackage{amsmath}
\usepackage{multirow}
\usepackage{dsfont}
\usepackage[shortlabels]{enumitem}
\usepackage{epstopdf}
\newcommand\BibTeX{{\rmfamily B\kern-.05em \textsc{i\kern-.025em b}\kern-.08em
T\kern-.1667em\lower.7ex\hbox{E}\kern-.125emX}}
\articletype{ORIGINAL ARTICLE}%
\received{<day> <Month>, <year>}
\revised{<day> <Month>, <year>}
\accepted{<day> <Month>, <year>}
\begin{document}
\title{Dynamic properties of the warm dense uniform electron gas with the qSTLS dielectric scheme}
\author{Panagiotis Tolias$^1$}
\author{Fotios Kalkavouras$^1$}
\author{Tobias Dornheim$^{2,3}$}
\author{Federico Lucco Castello$^{1}$}
\address{$^1$\orgdiv{Space and Plasma Physics}, \orgname{Royal Institute of Technology (KTH)}, \orgaddress{\state{Stockholm}, \country{Sweden}}}
\address{$^2$\orgname{Center for Advanced Systems Understanding (CASUS)}, \orgaddress{\state{G\"orlitz}, \country{Germany}}}
\address{$^3$\orgname{Helmholtz-Zentrum Dresden-Rossendorf (HZDR)}, \orgaddress{\state{Dresden}, \country{Germany}}}
\authormark{P. Tolias}
\corres{Panagiotis Tolias, Space and Plasma Physics, Royal Institute of Technology, Stockholm SE-100 44, Sweden. \email{tolias@kth.se}}
\abstract[Abstract]{The recently derived Fourier--Matsubara expansion of imaginary--time correlation functions comprises an exact result of linear response theory for finite-temperature quantum many-body systems. In its density--density version, the expansion facilitates systematic comparisons between quasi-exact \emph{ab initio} path integral Monte Carlo simulations and approximate dielectric formalism schemes at the level of the imaginary--time (density--density) correlation functions and the dynamic Matsubara local field corrections. On this theoretical basis, the dynamic properties of the quantum version of the Singwi--Tosi--Land--Sj\"olander scheme are analyzed for the paramagnetic warm dense uniform electron gas. The marginal improvement compared to the semi-classical version of the Singwi--Tosi--Land--Sj\"olander scheme is attributed to the weak Matsubara order dependence of the approximate dynamic Matsubara local field correction. The evaluation of the ideal density response function at the non-interacting occupation numbers is identified to constitute a general deficiency of the dielectric formalism, which calls for a reformulation in future works.}
\jnlcitation{\cname{%
\author{P. Tolias, F. Kalkavouras, T. Dornheim and F. Lucco Castello}} (\cyear{2024}),
\ctitle{Dynamic properties of the warm dense uniform electron gas with the qSTLS dielectric scheme}, \cjournal{Contributions to Plasma Physics}.}
\maketitle

\section{Introduction}\label{sec:intro}

Warm dense matter (WDM) denotes high energy density states characterized by high temperatures ($10^4-10^8\,$K), high pressures ($1-10^4\,$GBar) and beyond-solid densities ($10^{22}-10^{27}\,$cm$^{-3}$)\cite{GrazianiBook,RileyBook}. There has been growing interest in WDM due to its relevance to the early stages of intense laser-solid interactions and inertial confinement fusion\cite{HuPRL2010,KangMRE2020}, to the physics and evolution of dense astrophysical objects\cite{ChabrierAPJ2019,BeznogovPhysRep2021,SaumonPhysRep2022} as well as to the fabrication of novel materials under extreme conditions\cite{KrausNat2016,LazickiNat2021,SchusterJPD2022}. A pre-requisite for the understanding of WDM is the accurate description of the warm dense uniform electron gas (UEG), a fundamental homogeneous model system that does not require explicit treatment of the ionic component\cite{DornheimPhysRep2018,DornheimPoPRev2023}.

In the thermodynamically favorable unpolarized case of equal spin-up/-down electrons, the finite temperature UEG is fully characterized by two dimensionless parameters\cite{DornheimPhysRep2018,DornheimPoPRev2023}: the \emph{coupling parameter} $r_{\mathrm{s}}=d/a_{\mathrm{B}}$ with $d=(4\pi{n}/3)^{-1/3}$ the Wigner-Seitz radius, $a_{\mathrm{B}}=\hbar^2/(m_{\mathrm{e}}e^2)$ the Bohr radius and the \emph{degeneracy parameter} $\Theta=T/E_{\mathrm{F}}$ with $E_{\mathrm{F}}=\hbar^2k_{\mathrm{F}}^2/(2m)$ the Fermi energy with respect to the Fermi wavevector $k_{\mathrm{F}}=(3\pi^2n)^{1/3}$, $T$ the temperature in energy units. In the WDM regime, $r_{\mathrm{s}}\sim\Theta\sim1$; this lack of small parameters leads to difficulties in first-principles descriptions\cite{DornheimPhysRep2018,DornheimPoPRev2023,BonitzPoPRev2020}. Fortunately, in the past ten years, there have been numerous breakthroughs in quantum Monte Carlo simulations with the invention of entirely new techniques or the conception of novel variants of existing techniques that provide quasi-exact thermodynamic, structural and dynamic UEG results in previously inaccessible - owing to the fermion sign problem~\cite{TroyerPRL2005,DornheimPRE2019} - WDM regions\cite{SchoofCPP2011,DornheimNJP2015,MalonePRL2016,LeeJCP2021,DornheimJCP2020,YilmazJCP2020,XiongJCP2022,DornheimJCP2023}. These have led to a wealth of invaluable benchmark data that can be utilized to validate less computationally demanding approaches\cite{BonitzPoPRev2020}.

The dielectric formulation of the many body problem\cite{GiulianiVignale,NozieresINC1958,SingwiSSP1981} has undoubtedly emerged as one of the most popular, accurate and versatile microscopic frameworks for the description of the finite temperature or ground state UEG\cite{DornheimPhysRep2018,DornheimPoPRev2023,GiulianiVignale,NozieresINC1958,SingwiSSP1981,IchimaruPhysRep1987,IchimaruRevModPhys1993}. The self-consistent dielectric formalism combines exact results of linear density response theory\cite{GiulianiVignale,NozieresINC1958,SingwiSSP1981} with a closure stemming from approximate perturbative kinetic theory approaches\cite{SingwiPR1968,SingwiPR1970,IchimaruPRA1970,VashishtaPRB1972,PathakPRB1973,HasegawaJPSJ1975,UtsumiPRB1980,ToliasPRB2024}, approximate non-perturbative statistical mechanics approaches.\cite{MukhopadyayPRB1978,TanakaPRA1987,TanakaJCP2016,DornheimPRB2020,ToliasJCP2021,LuccoCastelloEPL2022,ToliasJCP2023} or parameterized quasi-exact quantum Monte Carlo results\cite{DornheimPRL2020,DornheimPRB2021}.

Regardless of the dielectric scheme, the self-consistent dielectric formalism has three ingredients. \emph{First}, in the polarization potential approach\cite{IchimaruBook}, a constitutive relation can be derived that expresses the density response function $\chi(\boldsymbol{k},\omega)$ in terms of the ideal (Lindhard) density response $\chi_0(\boldsymbol{k},\omega)$, the pair interaction $U(\boldsymbol{k})$ and the dynamic local field correction (LFC) $G(\boldsymbol{k},\omega)$ as
\begin{equation}
\chi(\boldsymbol{k},\omega)=\frac{\chi_0(\boldsymbol{k},\omega)}{1-U(\boldsymbol{k})\left[1-G(\boldsymbol{k},\omega)\right]\chi_0(\boldsymbol{k},\omega)}\,,\label{eq:densityresponseDLFC}
\end{equation}
where $U(\boldsymbol{k})=4\pi{e}^2/k^2$ for Coulomb interactions. \emph{Second}, for the finite temperature UEG, the zero frequency moment sum rule that connects the static structure factor (SSF) $S(\boldsymbol{k})$ with the dynamic structure factor (DSF) $S(\boldsymbol{k},\omega)$, the fluctuation--dissipation theorem that connects the imaginary part of the density response function with the DSF and the analytic continuation of the causal $\chi(\boldsymbol{k},\omega)$ to the complex frequency plane $\widetilde{\chi}(\boldsymbol{k},z)$ lead to a Matsubara series expression for the SFF that reads as\cite{TanakaJPSJ1986,TanakaPRA1985}
\begin{equation}
S(\boldsymbol{k})=-\frac{1}{{n}\beta}\displaystyle\sum_{l=-\infty}^{\infty}\widetilde{\chi}(\boldsymbol{k},\imath\omega_l)\,,\label{eq:MatsubaraSeries}
\end{equation}
with $\omega_l=2\pi{l}/(\beta\hbar)$ the bosonic Matsubara frequencies and $\beta$ the inverse temperature in energy units. The $\,\widetilde{}\,$ symbol over dynamic quantities signifies analytic continuation from the real frequency domain $\omega$ to the complex frequency domain $z$. \emph{Third}, the dynamic LFC is a complicated SSF functional of the form
\begin{equation}
G(\boldsymbol{k},\omega)\equiv{G}[S](\boldsymbol{k},\omega)\,.\label{eq:functionalclosure}
\end{equation}
The combination of Eqs.(\ref{eq:densityresponseDLFC},\ref{eq:MatsubaraSeries},\ref{eq:functionalclosure}) leads to a non-linear functional equation of the type
\begin{equation}
S(\boldsymbol{k})=-\frac{1}{{n}\beta}\displaystyle\sum_{l=-\infty}^{\infty}\frac{\widetilde{\chi}_0(\boldsymbol{k},\imath\omega_l)}{1-U(\boldsymbol{k})\{1-\widetilde{G}[S](\boldsymbol{k},\imath\omega_l)\}\widetilde{\chi}_0(\boldsymbol{k},\imath\omega_l)}\,,\label{eq:functionalclosurefull}
\end{equation}
that can be numerically solved for the SSF\cite{DornheimPhysRep2018,TanakaJPSJ1986}.

Dielectric schemes can be primarily categorized into static schemes that feature a static LFC, $G(\boldsymbol{k})$, and dynamic schemes that feature a dynamic LFC, $G(\boldsymbol{k},\omega)$. The random phase approximation (RPA) that features a zero LFC, $G(\boldsymbol{k},\omega)\equiv0$, deserves a special place in such classification. Most static dielectric schemes are semi-classical in the sense that they incorporate quantum effects only through the ideal Lindhard density response. The Singwi--Tosi--Land--Sj\"olander (STLS) scheme, that can be derived by imposing the eponymous closure condition in the equations of classical kinetic theory, constitutes the archetypal static dielectric theory\cite{SingwiPR1968}. On the other hand, dynamic dielectric schemes incorporate quantum effects beyond the RPA level. The Hasegawa--Shimizu scheme or quantum STLS (qSTLS), that can be derived by imposing the STLS closure condition in the equations of the quantum kinetic theory, constitutes the archetypal dynamic dielectric theory\cite{HasegawaJPSJ1975}.

Despite the numerous applications of different dielectric schemes to the finite temperature UEG\cite{TanakaJCP2016,DornheimPRB2020,ToliasJCP2021,LuccoCastelloEPL2022,ToliasJCP2023,KumarPRB2009,AroraEPJB2017}, the discussion is typically confined to the thermodynamic and structural properties. Dynamic properties are generally not investigated even for dynamic dielectric schemes. This is mainly due to the sparsity of quasi-exact quantum Monte Carlo results for the DSF\cite{DornheimPRL2018,GrothPRB2019,DornheimPRB2020b}; a direct consequence of the so-called analytic continuation (AC) problem. To be more specific, path integral Monte Carlo (PIMC) simulations provide direct access to the imaginary--time correlation function (ITCF) $F(\boldsymbol{k},\tau)$ of the finite temperature UEG\cite{CeperleyRMP1995}, which is defined as the wavenumber resolved density--density correlation function at the imaginary--time $t=-\imath\hbar\tau$ and is connected to the DSF through a two-sided Laplace transform
\begin{align}
F(\boldsymbol{k},\tau)=\int_{-\infty}^{+\infty}S(\boldsymbol{k},\omega)e^{-\hbar\omega\tau}d\omega\,.\label{eq:ITCFdef}
\end{align}
The inversion of this two-sided Laplace transform, which is required to convert the ITCF PIMC data to DSF data, is ill-conditioned with respect to the omnipresent Monte Carlo error bars, which comprises the AC problem\cite{JarrellPhysRep1996,ShaoPhysRep2023}. Given the uniqueness of two-sided Laplace transforms, a shift from doing physics in the real frequency domain of DSFs to doing physics in the imaginary--time domain of ITCFs has been recently advocated\cite{DornheimPoPRev2023,DornheimMRE2023,DornheimPTRSA2023,DornheimPRB2023}, which has already led to multiple breakthroughs in the analysis of X-ray Thomson scattering measurements\cite{DornheimNat2022,DornheimPoP2023,DornheimSciRep2024,dornheim2024modelfreerayleighweightxray,dornheim2024unravelingelectroniccorrelationswarm}. In particular, it turns out that the dielectric formalism is an ideal theoretical framework for dynamic predictions directly in the imaginary--time domain, benefitting from a recently-derived exact result of linear response theory which expresses the ITCF through the infinite series\cite{ToliasJCP2024}
\begin{align}
F(\boldsymbol{k},\tau)=-\frac{1}{n\beta}\sum_{l=-\infty}^{+\infty}\widetilde{\chi}(\boldsymbol{k},\imath\omega_l)e^{-\imath\hbar\omega_l\tau}\,.\label{eq:FourierMatsubaraSeries}
\end{align}
Since $F(\boldsymbol{k},0)=S(\boldsymbol{k})$, the Fourier-Matsubara series for the ITCF [see Eq.(\ref{eq:FourierMatsubaraSeries})] generalizes the Matsubara series for the SSF [see Eq.(\ref{eq:MatsubaraSeries})] to finite imaginary--times. In the dielectric formalism, the density response function at the imaginary Matsubara frequencies is automatically evaluated within the computational loop that solves the functional equation of Eq.(\ref{eq:functionalclosurefull}). Therefore, the Fourier-Matsubara series of Eq.(\ref{eq:FourierMatsubaraSeries}) implies a straightforward evaluation of the ITCF that can be directly compared with ITCF PIMC data. It is also noted that the inverted form of the Fourier-Matsubara series\cite{ToliasJCP2024}
\begin{align}
\widetilde{\chi}(\boldsymbol{k},\imath\omega_l)=-n\int_0^{\beta}F(\boldsymbol{k},\tau)\cos{(\hbar\omega_l\tau)}d\tau\,.\label{eq:invDRFFourierMatsubaraSeries}
\end{align}
implies that the dynamic Matsubara density responses can be directly extracted from PIMC simulations\cite{DornheimPRB2024,DornheimEPL2024,MoldabekovPPNP2025,dornheim2024shortwavelengthlimitdynamic}. Finally, the analytically continued version of the constitutive relation of the polarization potential approach [see Eq.(\ref{eq:densityresponseDLFC})], but now solved for the dynamic Matsubara LFC, 
\begin{align}
\widetilde{G}(\boldsymbol{k},\imath\omega_l)=1-\frac{1}{U(\boldsymbol{k})}\left[\frac{1}{\widetilde{\chi}_0(\boldsymbol{k},\imath\omega_l)}-\frac{1}{\widetilde{\chi}(\boldsymbol{k},\imath\omega_l)}\right]\,,\label{eq:invLFCFourierMatsubaraSeries}
\end{align}
implies that also the dynamic Matsubara LFCs can be directly extracted from PIMC simulations\cite{DornheimPRB2024,DornheimEPL2024,dornheim2024shortwavelengthlimitdynamic}.

In this work, we report the first dielectric formalism investigation of the dynamic density--density correlation functions of the warm dense UEG in the imaginary--time domain, carried out on the basis of the Fourier-Matsubara expansion. The quantum version of the STLS scheme is considered, since it constitutes the prototypical dynamic dielectric theory. After a systematic comparison with benchmark data from existing PIMC simulations, conclusions are drawn about the specific limitations of the qSTLS scheme and the general limitations of the standard dielectric formalism in the description of the dynamic properties of the warm dense UEG.

\section{The qSTLS dielectric scheme}\label{sec:background}

\subsection{Theoretical aspects of the qSTLS scheme}\label{sec:qSTLStheoretical}

The qSTLS scheme constitutes the fully quantum version of the semi-classical STLS scheme. In the STLS scheme\cite{SingwiPR1968,TanakaJPSJ1986}, the classical BBGKY hierarchy of s-reduced distribution functions is truncated at its $s=1$ member with the ansatz $f_2(\boldsymbol{r},\boldsymbol{p},\boldsymbol{r}^{\prime},\boldsymbol{p}^{\prime},t)=f(\boldsymbol{r},\boldsymbol{p},t)f(\boldsymbol{r}^{\prime},\boldsymbol{p}^{\prime},t)g(\boldsymbol{r}-\boldsymbol{r}^{\prime})$ where $g(\boldsymbol{r})$ is the equilibrium pair correlation function. As a consequence, quantum effects enter only through the Lindhard density response and the LFC is static. In the qSTLS scheme\cite{HasegawaJPSJ1975,HolasPRB1987,SchwengPRB1993}, the quantum BBGKY hierarchy of s-reduced distribution functions is truncated at its $s=1$ member with the same ansatz $f_2(\boldsymbol{r},\boldsymbol{p},\boldsymbol{r}^{\prime},\boldsymbol{p}^{\prime},t)=f(\boldsymbol{r},\boldsymbol{p},t)f(\boldsymbol{r}^{\prime},\boldsymbol{p}^{\prime},t)g(\boldsymbol{r}-\boldsymbol{r}^{\prime})$, which leads to a more complete treatment of quantum effects and to a dynamic LFC. For completeness, let us first derive the dynamic LFC functional of the qSTLS scheme starting from the $s=1$ member of the quantum BBGKY hierarchy in the Wigner representation. In what follows, $U_{\mathrm{ext}}(\boldsymbol{r},t)$ is the external potential energy term and $U(\boldsymbol{r})$ is the pair interaction energy.
\begin{align*}
\left\{\frac{\partial}{\partial{t}}+\boldsymbol{v}\cdot\frac{\partial}{\partial\boldsymbol{r}}\right\}&f(\boldsymbol{r},\boldsymbol{p},t)=-\frac{1}{\imath{\hbar}}\int\,\frac{d^3\lambda{d}^3\bar{p}}{(2\pi)^3}\exp{\left[\imath\left(\boldsymbol{p}-\boldsymbol{\bar{p}}\right)\cdot\boldsymbol{\lambda}\right]}\left[U_{\mathrm{ext}}\left(\boldsymbol{r}+\frac{\hbar}{2}\boldsymbol{\lambda},t\right)-U_{\mathrm{ext}}\left(\boldsymbol{r}-\frac{\hbar}{2}\boldsymbol{\lambda},t\right)\right]f(\boldsymbol{r},\boldsymbol{\bar{p}},t)\nonumber\\&-\frac{1}{\imath{\hbar}}\int\,\frac{d^3\lambda{d}^3\bar{p}}{(2\pi)^3}d^3r^{\prime}d^3p^{\prime}\exp{\left[\imath\left(\boldsymbol{p}-\boldsymbol{\bar{p}}\right)\cdot\boldsymbol{\lambda}\right]}\left[U\left(\boldsymbol{r}-\boldsymbol{r}^{\prime}+\frac{\hbar}{2}\boldsymbol{\lambda}\right)-U\left(\boldsymbol{r}-\boldsymbol{r}^{\prime}-\frac{\hbar}{2}\boldsymbol{\lambda}\right)\right]f_2(\boldsymbol{r},\bar{\boldsymbol{p}},\boldsymbol{r}^{\prime},\boldsymbol{p}^{\prime},t)\,.\nonumber
\end{align*}
Application of an external potential energy perturbation $\delta{U}_{\mathrm{ext}}(\boldsymbol{r},t)$ to the equilibrium system, substitution of the STLS condition, perturbation of the one-particle distribution function around the equilibrium $f(\boldsymbol{r},\boldsymbol{p},t)\simeq{f}_0(\boldsymbol{p})+\delta{f}(\boldsymbol{r},\boldsymbol{p},t)$, linearization with respect to the small perturbation strength, use of spatiotemporal Fourier transforms, consideration of the adiabatic switching of the perturbation, use of $S(\boldsymbol{k})=1+nH(\boldsymbol{k})$ with $H(\boldsymbol{k})$ the Fourier transform of the total correlation function $h(\boldsymbol{r})=g(\boldsymbol{r})-1$ yield
\begin{align*}
\delta{f}(\boldsymbol{k},\boldsymbol{p},\omega)&=-\frac{1}{\hbar}\frac{f_0\left(\boldsymbol{p}+\frac{\hbar}{2}\boldsymbol{k}\right)-f_0\left(\boldsymbol{p}-\frac{\hbar}{2}\boldsymbol{k}\right)}{\omega-\boldsymbol{k}\cdot\boldsymbol{v}+\imath0}\delta{U}_{\mathrm{ext}}(\boldsymbol{k},\omega)-\frac{1}{\hbar}\frac{f_0\left(\boldsymbol{p}+\frac{\hbar}{2}\boldsymbol{k}\right)-f_0\left(\boldsymbol{p}-\frac{\hbar}{2}\boldsymbol{k}\right)}{\omega-\boldsymbol{k}\cdot\boldsymbol{v}+\imath0}U\left(\boldsymbol{k}\right)\delta{n}(\boldsymbol{k},\omega)\nonumber\\&\,\,\,\,\,\,-\frac{1}{\hbar}\left\{\frac{1}{n}\int\,\frac{d^3k^{\prime}}{(2\pi)^3}\frac{f_0\left(\boldsymbol{p}+\frac{\hbar}{2}\boldsymbol{k}^{\prime}\right)-f_0\left(\boldsymbol{p}-\frac{\hbar}{2}\boldsymbol{k}^{\prime}\right)}{\omega-\boldsymbol{k}\cdot\boldsymbol{v}+\imath0}U\left(\boldsymbol{k}^{\prime}\right)\left[S(\boldsymbol{k}-\boldsymbol{k}^{\prime})-1\right]\right\}\delta{n}(\boldsymbol{k},\omega)\,.\nonumber
\end{align*}
Integration over the momenta to obtain the density perturbation, solution of the resulting linear equation with respect to the density perturbation $\delta{n}(\boldsymbol{k},\omega)$, introduction of the ideal Lindhard density response $\chi_0(\boldsymbol{k},\omega)$
\begin{align}
\chi_0(\boldsymbol{k},\omega)=-\frac{2}{\hbar}\int\frac{d^3q}{(2\pi)^3}\frac{f_0\left(\boldsymbol{q}+\frac{1}{2}\boldsymbol{k}\right)-f_0\left(\boldsymbol{q}-\frac{1}{2}\boldsymbol{k}\right)}{\omega-\frac{\hbar}{m}\boldsymbol{k}\cdot\boldsymbol{q}+\imath0}\,,\label{eq:Lindhardgeneral}
\end{align}
introduction of the three-argument ideal density response $\chi_0(\boldsymbol{k},\boldsymbol{k}^{\prime},\omega)$, which obeys $\chi_0(\boldsymbol{k},\boldsymbol{k},\omega)\equiv\chi_0(\boldsymbol{k},\omega)$,
\begin{align}
\chi_0(\boldsymbol{k},\boldsymbol{k}^{\prime}\omega)=-\frac{2}{\hbar}\int\frac{d^3q}{(2\pi)^3}\frac{f_0\left(\boldsymbol{q}+\frac{1}{2}\boldsymbol{k}^{\prime}\right)-f_0\left(\boldsymbol{q}-\frac{1}{2}\boldsymbol{k}^{\prime}\right)}{\omega-\frac{\hbar}{m}\boldsymbol{k}\cdot\boldsymbol{q}+\imath0}\,,\label{eq:Lindhard3argumentgeneral}
\end{align}
and use of the functional derivative definition $\chi(\boldsymbol{k},\omega)=\delta{n}(\boldsymbol{k},\omega)/\delta{U}_{\mathrm{ext}}(\boldsymbol{k},\omega)$ yield the qSTLS density response function
\begin{align*}
\chi(\boldsymbol{k},\omega)=\frac{\chi_0(\boldsymbol{k},\omega)}{1-U\left(\boldsymbol{k}\right)\left\{1+\frac{1}{n}\displaystyle\int\frac{d^3k^{\prime}}{(2\pi)^3}\frac{U\left(\boldsymbol{k}^{\prime}\right)}{U\left(\boldsymbol{k}\right)}\frac{\chi_0(\boldsymbol{k},\boldsymbol{k}^{\prime},\omega)}{\chi_{0}(\boldsymbol{k},\omega)}\left[S(\boldsymbol{k}-\boldsymbol{k}^{\prime})-1\right]\right\}\chi^{0}_{\boldsymbol{k},\omega}}\,.\nonumber
\end{align*}
Comparison with the constitutive relation [see Eq.(\ref{eq:densityresponseDLFC})] that serves as the general definition of the LFC, ultimately leads to the qSTLS functional for the dynamic LFC
\begin{align}
G_{\mathrm{qSTLS}}(\boldsymbol{k},\omega)&=-\frac{1}{n}\displaystyle\int\frac{d^3k^{\prime}}{(2\pi)^3}\frac{U\left(\boldsymbol{k}^{\prime}\right)}{U\left(\boldsymbol{k}\right)}\frac{\chi_0(\boldsymbol{k},\boldsymbol{k}^{\prime},\omega)}{\chi_{0}(\boldsymbol{k},\omega)}\left[S(\boldsymbol{k}-\boldsymbol{k}^{\prime})-1\right]\,.\label{eq:qSTLSfunctionalLFC}
\end{align}
For computational reasons, it is preferable to write the qSTLS density response function as
\begin{align}
\chi(\boldsymbol{k},\omega)=\frac{\chi_0(\boldsymbol{k},\omega)}{1-U\left(\boldsymbol{k}\right)\chi_0(\boldsymbol{k},\omega)+U\left(\boldsymbol{k}\right)I_{\mathrm{qSTLS}}(\boldsymbol{k},\omega)}\,,\label{eq:constitutivedynamic}
\end{align}
where the qSTLS auxiliary response is given by $I_{\mathrm{qSTLS}}(\boldsymbol{k},\omega)=\chi_0(\boldsymbol{k},\omega)G_{\mathrm{qSTLS}}(\boldsymbol{k},\omega)$ or
\begin{align}
I_{\mathrm{qSTLS}}(\boldsymbol{k},\omega)=-\frac{1}{n}\displaystyle\int\frac{d^3k^{\prime}}{(2\pi)^3}\frac{U\left(\boldsymbol{k}^{\prime}\right)}{U\left(\boldsymbol{k}\right)}\chi_0(\boldsymbol{k},\boldsymbol{k}^{\prime},\omega)\left[S(\boldsymbol{k}-\boldsymbol{k}^{\prime})-1\right]\,.\label{eq:qSTLSfunctionalLFCaux}
\end{align}
It is noted that the STLS functional for the static LFC is given by
\begin{align}
G_{\mathrm{STLS}}(\boldsymbol{k})&=-\frac{1}{n}\displaystyle\int\frac{d^3k^{\prime}}{(2\pi)^3}\frac{U\left(\boldsymbol{k}^{\prime}\right)}{U\left(\boldsymbol{k}\right)}\frac{\boldsymbol{k}\cdot\boldsymbol{k}^{\prime}}{k^2}\left[S(\boldsymbol{k}-\boldsymbol{k}^{\prime})-1\right]\,.\label{eq:STLSfunctionalLFC}
\end{align}
Thus, irrespective of the pair interaction (kept arbitrary thus far), the qSTLS functional can be obtained from the STLS functional after the substitution $\boldsymbol{k}\cdot\boldsymbol{k}^{\prime}/k^2\to\chi_0(\boldsymbol{k},\boldsymbol{k}^{\prime},\omega)/\chi_{0}(\boldsymbol{k},\omega)$\cite{ToliasJCP2023}. Very recently\cite{ToliasJCP2023}, this correspondence rule was successfully employed for the rapid quantization of semi-classical dielectric schemes tailor-made for the strongly coupled regime\cite{ToliasJCP2021,LuccoCastelloEPL2022}.

\subsection{Computational aspects of the qSTLS scheme}\label{sec:qSTLSnumerical}

In what follows, we shall cast the closed set of equations for the qSTLS scheme into a numerically convenient form. First, after the utilization of spherical coordinates and the substitution for the Fermi-Dirac distribution, the Lindhard density response [see Eq.(\ref{eq:Lindhardgeneral}) but now evaluated at the bosonic Matsubara frequencies $\imath\omega_l=2\pi\imath{l}/\beta\hbar$] becomes
\begin{align*}
\widetilde{\chi}_0(\boldsymbol{k},\imath\omega_l)&=-\frac{2}{k}\frac{m}{\hbar^2}\int_0^{\infty}\frac{dq}{(2\pi)^2}\frac{q}{\exp{\left(\frac{\hbar^2q^2}{2mT}-\frac{\mu}{T}\right)}+1}\ln{\left[\frac{\left(k^2+2qk\right)^2+\left(\frac{4\pi{l}mT}{\hbar^2}\right)^2}{\left(k^2-2qk\right)^2+\left(\frac{4\pi{l}mT}{\hbar^2}\right)^2}\right]}\,,\nonumber\\
\widetilde{\chi}_0(\boldsymbol{k},0)&=-\frac{2}{k}\frac{1}{T}\int_0^{\infty}\frac{dq}{(2\pi)^2}\left[\left(q^2-\frac{k^2}{4}\right)\ln{\left|\frac{2q+k}{2q-k}\right|}+qk\right]\frac{q\exp{\left(\frac{\hbar^2q^2}{2mT}-\frac{\mu}{T}\right)}}{\left[\exp{\left(\frac{\hbar^2q^2}{2mT}-\frac{\mu}{T}\right)}+1\right]^2}\,,\nonumber
\end{align*}
where the ground state logarithmic singularity ($l=0$) at $q=k/2$ has been removed by partial integration. After the introduction of the normalizations $x=k/k_{\mathrm{F}}$, $y=q/k_{\mathrm{F}}$ and $\bar{\mu}=\mu/T$, the normalized ideal Matsubara density response becomes
\begin{align}
\frac{\widetilde{\chi}_0(x,\imath\omega_l)}{n\beta}&=-\frac{3}{4}\frac{\Theta}{x}\int_0^{\infty}dy\frac{y}{\exp{\left(\frac{y^2}{\Theta}-\bar{\mu}\right)}+1}\ln{\left[\frac{\left(x^2+2xy\right)^2+\left(2\pi{l}{\Theta}\right)^2}{\left(x^2-2xy\right)^2+\left(2\pi{l}\Theta\right)^2}\right]}\,,\label{eq:LindhardMatsuNorm}\\
\frac{\widetilde{\chi}_0(x,0)}{n\beta}&=-\frac{3}{2}\frac{1}{x}\int_0^{\infty}dy\left[\left(y^2-\frac{x^2}{4}\right)\ln{\left|\frac{2y+x}{2y-x}\right|}+xy\right]\frac{y\exp{\left(\frac{y^2}{\Theta}-\bar{\mu}\right)}}{\left[\exp{\left(\frac{y^2}{\Theta}-\bar{\mu}\right)}+1\right]^2}\,.\label{eq:LindhardMatsu0Norm}
\end{align}
Moreover, after the utilization of spherical coordinates and the substitution for the Fermi-Dirac distribution, the three-argument ideal density response [see Eq.(\ref{eq:Lindhard3argumentgeneral}) but now evaluated at the bosonic Matsubara frequencies $\imath\omega_l=2\pi\imath{l}/\beta\hbar$] becomes
\begin{align*}
\widetilde{\chi}_0(\boldsymbol{k},\boldsymbol{k}^{\prime},\imath\omega_l)&=-\frac{2}{k}\frac{m}{\hbar^2}\int_0^{\infty}\frac{dq}{(2\pi)^2}\frac{q}{\exp{\left(\frac{\hbar^2q^2}{2mT}-\frac{\mu}{T}\right)}+1}\ln{\left[\frac{\left(\boldsymbol{k}\cdot\boldsymbol{k}^{\prime}+2qk\right)^2+\left(\frac{4{\pi}lmT}{\hbar^2}\right)^2}{\left(\boldsymbol{k}\cdot\boldsymbol{k}^{\prime}-2qk\right)^2+\left(\frac{4{\pi}lmT}{\hbar^2}\right)^2}\right]}\,,\nonumber\\
\widetilde{\chi}_0(\boldsymbol{k},\boldsymbol{k}^{\prime},0)&=-\frac{2}{k}\frac{1}{T}\int_0^{\infty}\frac{dq}{(2\pi)^2}\left\{\left[q^2-\frac{(\boldsymbol{k}\cdot\boldsymbol{k}^{\prime})^2}{4k^2}\right]\ln{\left|\frac{\boldsymbol{k}\cdot\boldsymbol{k}^{\prime}+2kq}{\boldsymbol{k}\cdot\boldsymbol{k}^{\prime}-2kq}\right|}+q\frac{\boldsymbol{k}\cdot\boldsymbol{k}^{\prime}}{k}\right\}\frac{q\exp{\left(\frac{\hbar^2q^2}{2mT}-\frac{\mu}{T}\right)}}{\left[\exp{\left(\frac{\hbar^2q^2}{2mT}-\frac{\mu}{T}\right)}+1\right]^2}\,,\nonumber
\end{align*}
where the ground state logarithmic singularity ($l=0$) at $\boldsymbol{k}\cdot\boldsymbol{k}^{\prime}=2qk$ has again been removed by partial integration. After the introduction of the same normalizations, the normalized three-argument ideal Matsubara density response becomes
\begin{align*}
\frac{\widetilde{\chi}_0(\boldsymbol{x},\boldsymbol{x}^{\prime},\imath\omega_l)}{n\beta}&=-\frac{3}{4}\frac{\Theta}{x}\int_0^{\infty}dy\frac{y}{\exp{\left(\frac{y^2}{\Theta}-\bar{\mu}\right)}+1}\ln{\left[\frac{\left(\boldsymbol{x}\cdot\boldsymbol{x}^{\prime}+2xy\right)^2+\left(2{\pi}l\Theta\right)^2}{\left(\boldsymbol{x}\cdot\boldsymbol{x}^{\prime}-2xy\right)^2+\left(2{\pi}l\Theta\right)^2}\right]}\,,\nonumber\\
\frac{\widetilde{\chi}_0(\boldsymbol{x},\boldsymbol{x}^{\prime},0)}{n\beta}&=-\frac{3}{2}\frac{1}{x}\int_0^{\infty}dy\left\{\left[y^2-\frac{(\boldsymbol{x}\cdot\boldsymbol{x}^{\prime})^2}{4x^2}\right]\ln{\left|\frac{\boldsymbol{x}\cdot\boldsymbol{x}^{\prime}+2xy}{\boldsymbol{x}\cdot\boldsymbol{x}^{\prime}-2xy}\right|}+y\frac{\boldsymbol{x}\cdot\boldsymbol{x}^{\prime}}{x}\right\}\frac{y\exp{\left(\frac{y^2}{\Theta}-\bar{\mu}\right)}}{\left[\exp{\left(\frac{y^2}{\Theta}-\bar{\mu}\right)}+1\right]^2}\,.\nonumber
\end{align*}
Furthermore, after the utilization of spherical coordinates, the substitution for the Fourier transformed Coulomb pair potential energy, the substitution for the three-argument ideal density response and the standard normalizations, the normalized qSTLS auxiliary Matsubara response [see Eq.(\ref{eq:qSTLSfunctionalLFCaux}) but now evaluated at the bosonic Matsubara frequencies $\imath\omega_l=2\pi\imath{l}/\beta\hbar$] becomes
\begin{align}
&\frac{\widetilde{I}_{\mathrm{qSTLS}}(x,\imath\omega_l)}{n\beta}=\frac{9}{16}\Theta\int_0^{\infty}w\left[S(w)-1\right]dw\int_0^{\infty}\frac{ydy}{\exp{\left(\frac{y^2}{\Theta}-\bar{\mu}\right)}+1}\int_{x^2-xw}^{x^2+xw}\frac{dt}{2t+w^2-x^2}\ln{\left[\frac{\left(2xy+t\right)^2+\left(2{\pi}l\Theta\right)^2}{\left(2xy-t\right)^2+\left(2{\pi}l\Theta\right)^2}\right]}\,,\label{eq:auxqSTLSMatsuNorm}\\
&\frac{\widetilde{I}_{\mathrm{qSTLS}}(x,0)}{n\beta}=\frac{9}{8}\int_0^{\infty}w\left[S(w)-1\right]dw\int_0^{\infty}\frac{y\exp{\left(\frac{y^2}{\Theta}-\bar{\mu}\right)}dy}{\left[\exp{\left(\frac{y^2}{\Theta}-\bar{\mu}\right)}+1\right]^2}\int_{x^2-xw}^{x^2+xw}\frac{dt}{2t+w^2-x^2}\left[\left(y^2-\frac{t^2}{4x^2}\right)\ln{\left|\frac{t+2xy}{t-2xy}\right|}+\frac{y}{x}t\right]\,.\label{eq:auxqSTLSMatsu0Norm}
\end{align}
Finally, following the seminal work of Tanaka \& Ichimaru for the STLS scheme at finite temperatures\cite{TanakaJPSJ1986}, we introduce the auxiliary complex functions $\Phi(x,l)=-(2E_{\mathrm{F}}/3n)\widetilde{\chi}_0(x,\imath\omega_l)$ and $\Psi(x,l)=-(2E_{\mathrm{F}}/3n)\widetilde{I}_{\mathrm{qSTLS}}(x,\imath\omega_l)$. Overall, the closed normalized set of equations for the qSTLS scheme comprises the normalization condition of the Fermi distribution [see Eq.(\ref{qSTLSfinal1})], the normalized ideal Matsubara density response expressed through the auxiliary complex function $\Phi(x,l)$ including the static limit [see Eqs.(\ref{qSTLSfinal2},\ref{qSTLSfinal3})], the normalized qSTLS auxiliary Matsubara response expressed through the auxiliary complex function $\Psi(x,l)$ including the static limit [see Eqs.(\ref{qSTLSfinal4},\ref{qSTLSfinal5})] and the infinite Matsubara summation expression for the SSF [see Eq.(\ref{qSTLSfinal6})].
\begin{align}
&\int_0^{\infty}\frac{\sqrt{z}dz}{\exp{\left(z-\bar{\mu}\right)}+1}=\frac{2}{3}\Theta^{-3/2}\,,\label{qSTLSfinal1}\\
&\Phi(x,l)=\frac{1}{2x}\int_0^{\infty}dy\frac{y}{\exp{\left(\frac{y^2}{\Theta}-\bar{\mu}\right)}+1}\ln{\left[\frac{\left(x^2+2xy\right)^2+\left(2\pi{l}{\Theta}\right)^2}{\left(x^2-2xy\right)^2+\left(2\pi{l}\Theta\right)^2}\right]}\,,\label{qSTLSfinal2}\\
&\Phi(x,0)=\frac{1}{\Theta{x}}\int_0^{\infty}dy\left[\left(y^2-\frac{x^2}{4}\right)\ln{\left|\frac{2y+x}{2y-x}\right|}+xy\right]\frac{y\exp{\left(\frac{y^2}{\Theta}-\bar{\mu}\right)}}{\left[\exp{\left(\frac{y^2}{\Theta}-\bar{\mu}\right)}+1\right]^2}\,,\label{qSTLSfinal3}\\
&\Psi(x,l)=-\frac{3}{8}\int_0^{\infty}w\left[S(w)-1\right]dw\int_0^{\infty}\frac{ydy}{\exp{\left(\frac{y^2}{\Theta}-\bar{\mu}\right)}+1}\int_{x^2-xw}^{x^2+xw}\frac{dt}{2t+w^2-x^2}\ln{\left[\frac{\left(2xy+t\right)^2+\left(2{\pi}l\Theta\right)^2}{\left(2xy-t\right)^2+\left(2{\pi}l\Theta\right)^2}\right]}\,,\label{qSTLSfinal4}\\
&\Psi(x,0)=-\frac{3}{4\Theta}\int_0^{\infty}w\left[S(w)-1\right]dw\int_0^{\infty}\frac{y\exp{\left(\frac{y^2}{\Theta}-\bar{\mu}\right)}dy}{\left[\exp{\left(\frac{y^2}{\Theta}-\bar{\mu}\right)}+1\right]^2}\int_{x^2-xw}^{x^2+xw}\frac{dt}{2t+w^2-x^2}\left[\left(y^2-\frac{t^2}{4x^2}\right)\ln{\left|\frac{t+2xy}{t-2xy}\right|}+\frac{y}{x}t\right]\,,\label{qSTLSfinal5}\\
&S(x)=\frac{3}{2}\Theta\displaystyle\sum_{l=-\infty}^{\infty}\frac{\Phi(x,l)}{1+\displaystyle\frac{4}{\pi}\lambda{r}_{\mathrm{s}}\frac{1}{x^2}\left[\Phi(x,l)-\Psi(x,l)\right]}\,.\label{qSTLSfinal6}
\end{align}
In order to speed up the Matsubara series convergence, again following the seminal work of Tanaka \& Ichimaru for the STLS scheme at finite temperatures\cite{TanakaJPSJ1986}, the non-interacting (Hartree-Fock) SSF, has been split up from the infinite series, as in
\begin{align}
S(x)=S_{\mathrm{HF}}(x)-\frac{6}{\pi}\lambda{r}_{\mathrm{s}}\Theta\frac{1}{x^2}\displaystyle\sum_{l=-\infty}^{\infty}\frac{\Phi(x,l)\left[\Phi(x,l)-\Psi(x,l)\right]}{1+\displaystyle\frac{4}{\pi}\lambda{r}_{\mathrm{s}}\frac{1}{x^2}\left[\Phi(x,l)-\Psi(x,l)\right]}\,.\label{qSTLSfinal7}
\end{align}
In the above, the numerical constant $\lambda=[4/(9\pi)]^{1/3}$ is defined by $\lambda=1/(k_{\mathrm{F}}d)$. The non-interacting SSF can be evaluated through the single integral\cite{TanakaJPSJ1986,tolias2024noninteractingUEG}
\begin{align}
S_{\mathrm{HF}}(x)=1-\frac{3\Theta}{4x}\int_{0}^{+\infty}\frac{y}{\exp{\left(\frac{y^2}{\Theta}-\bar{\mu}\right)}+1}\ln{\left|\frac{1+\exp{\left[\bar{\mu}-\frac{(y-x)^2}{\Theta}\right]}}{1+\exp{\left[\bar{\mu}-\frac{(y+x)^2}{\Theta}\right]}}\right|}dy\,.\label{qSTLSfinal8}
\end{align}
For each UEG state point, the iteration cycle proceeds as follows. The reduced chemical potential is calculated from Eq.(\ref{qSTLSfinal1}) with the Brent-Dekker hybrid root-finding algorithm and the non-interacting SSF is computed from Eq.(\ref{qSTLSfinal8}). The normalized Lindhard Matsubara density responses $\Phi(x,l)$ are computed from Eqs.(\ref{qSTLSfinal2},\ref{qSTLSfinal3}) for the Matsubara orders $0\leq{l}\leq500$. The normalized qSTLS auxiliary Matsubara responses $\Psi(x,l)$ are computed from Eqs.(\ref{qSTLSfinal4},\ref{qSTLSfinal5}) also for the Matsubara orders $0\leq{l}\leq500$ with the RPA SSF as the initial guess. The SSF guess is updated from Eq.(\ref{qSTLSfinal7}). The last two steps are repeated until the absolute relative difference between two successive qSTLS SSF evaluations is smaller than $10^{-5}$ for all the grid points. The improper integrals are handled with the doubly adaptive general-purpose quadrature routine CQUAD of the GSL library; a $0.1$ wavenumber grid resolution and a $50$ upper cutoff wavenumber are employed.

\subsection{Post-processing aspects of the qSTLS scheme}\label{sec:qSTLSpostprocessing}

The dielectric formalism yields direct access to the static structure factor, dynamic Matsubara density response and dynamic Matsubara local field correction. Here, we shall briefly present the extra steps required to calculate the interaction energy, pair correlation function (PCF), imaginary--time correlation function and thermal structure factor. The thermodynamic route is opened up by the interaction energy (per electron) expression $u_{\mathrm{int}}=(1/2)\int\,U(k)\left[S(k)-1\right][d^3k/(2\pi)^3]$. In normalized units, the expression becomes
\begin{align}
\widetilde{u}_{\mathrm{int}}(r_{\mathrm{s}},\Theta)&=\frac{1}{\pi\lambda{r}_{\mathrm{s}}}\int_0^{\infty}\left[S(x;r_{\mathrm{s}},\Theta)-1\right]dx\,,\label{qSTLSpost1}
\end{align}
where the $\,\widetilde{}\,$ symbol signifies that the interaction energy is expressed in Hartree energy units, $e^2/a_{\mathrm{B}}$. The PCF is evaluated from the SSF via the Fourier inversion of $S(\boldsymbol{k})=1+nH(\boldsymbol{k})$. In normalized units, the expression becomes
\begin{align}
g(x)=1+\frac{3}{2x}\int_0^{\infty}y\sin{(xy)}\left[S(y)-1\right]dy\,.\label{qSTLSpost2}
\end{align}
The ITCF is evaluated from its Fourier--Matsubara series [see Eq.(\ref{eq:FourierMatsubaraSeries})]. Similar to the SSF case, the convergence is accelerated by splitting up the non-interacting ITCF from the infinite series. In normalized units, with $\tau^{\star}={T}\tau$, the expression becomes\cite{tolias2024noninteractingUEG}
\begin{align}
F(x,\tau^{\star})=F_{\mathrm{HF}}(x,\tau^{\star})-\frac{6}{\pi}\lambda{r}_{\mathrm{s}}\Theta\frac{1}{x^2}\displaystyle\sum_{l=-\infty}^{\infty}\frac{\Phi(x,l)\left[\Phi(x,l)-\Psi(x,l)\right]}{1+\displaystyle\frac{4}{\pi}\lambda{r}_{\mathrm{s}}\frac{1}{x^2}\left[\Phi(x,l)-\Psi(x,l)\right]}\cos{(2\pi{l}\tau^{\star})}\,.\label{qSTLSpost3}
\end{align}
where the non-interacting ITCF can be evaluated through the single integral\cite{tolias2024noninteractingUEG}
\begin{align}
F_{\mathrm{HF}}(x,\tau^{\star})&=+\frac{3\Theta}{8}\int_{0}^{+\infty}\frac{\cosh{\left[\frac{xy}{\Theta}(\tau^{\star}-1/2)\right]}}{\sinh{\left(\frac{xy}{2\Theta}\right)}}\ln{\left\{\frac{1+\exp{\left[\bar{\mu}-\frac{(x-y)^2}{4\Theta}\right]}}{1+\exp{\left[\bar{\mu}-\frac{(x+y)^2}{4\Theta}\right]}}\right\}}dy\,.\label{qSTLSpost4}
\end{align}
The TSF is evaluated by setting $\tau=\beta/2$ to the Fourier-Matsubara series [see Eq.(\ref{eq:FourierMatsubaraSeries})]. As usual, the convergence is accelerated by splitting up the non-interacting TSF from the infinite series. In normalized units, the expression becomes\cite{tolias2024noninteractingUEG}
\begin{align}
S^{\beta/2}(x)=S^{\beta/2}_{\mathrm{HF}}(x)-\frac{6}{\pi}\lambda{r}_{\mathrm{s}}\Theta\frac{1}{x^2}\displaystyle\sum_{l=-\infty}^{\infty}(-1)^l\frac{\Phi(x,l)\left[\Phi(x,l)-\Psi(x,l)\right]}{1+\displaystyle\frac{4}{\pi}\lambda{r}_{\mathrm{s}}\frac{1}{x^2}\left[\Phi(x,l)-\Psi(x,l)\right]}\,.\label{qSTLSpost5}
\end{align}
where the non-interacting TSF can be evaluated through the single integral\cite{tolias2024noninteractingUEG}
\begin{align}
S_{\mathrm{HF}}^{\beta/2}(x)&=\frac{3}{2}\frac{\Theta}{x}\int_{-\infty}^{\infty}dy\frac{y}{\exp{\left(\frac{y^2}{\Theta}-\bar{\mu}\right)}+1}\ln{\left|\frac{1-\exp{\left(-\frac{x^2+2xy}{2\Theta}\right)}}{1+\exp{\left(-\frac{x^2+2xy}{2\Theta}\right)}}\right|}\,.\label{qSTLSpost6}
\end{align}

\subsection{Limiting behavior of the qSTLS dynamic Matsubara local field correction}\label{sec:qSTLSasymptotics}

In what follows, we examine the limiting behavior of the qSTLS dynamic Matsubara LFC $\widetilde{G}_{\mathrm{qSTLS}}(x,l)$ with respect to the wavenumber (at zero and infinity) and with respect to the Matsubara frequency order (at zero and infinity). 

\textbf{\emph{The long-wavelength limit}}. By dividing the long-wavelength limit of the normalized qSTLS auxiliary Matsubara response [see Eq.(\ref{eq:longwavelengthQSTLS}) of Appendix B] with the long-wavelength limit of the normalized ideal Matsubara density response [see Eq.(\ref{eq:longwavelengthLindhard}) of Appendix A], we directly obtain the long-wavelength limit of the qSTLS dynamic Matsubara LFC
\begin{align}
\displaystyle\lim_{x\to0}\widetilde{G}_{\mathrm{qSTLS}}(x,l)=-\frac{1}{2}\pi\lambda{r}_{\mathrm{s}}\widetilde{u}_{\mathrm{int}}x^2\,,\label{eq:longwavelengthQSTLSLFC}
\end{align}
where $\widetilde{u}_{\mathrm{int}}=[1/(\pi\lambda{r}_{\mathrm{s}})]\int_0^{\infty}\left[S(w)-1\right]dw$ is the qSTLS interaction energy in Hartree units. It is not surprising that the long-wavelength limit can be expressed through a thermodynamic quantity. It is pointed out that the long-wavelength limit of the qSTLS dynamic Matsubara LFC does not depend on the order of the Matsubara frequency. This result was originally derived by Schweng \& B\"ohm\cite{SchwengPRB1993}; note that the factor of two difference originates from their use of Rydberg energy units.

\textbf{\emph{The short-wavelength limit}}. The ratio of the short-wavelength limit of the normalized qSTLS auxiliary Matsubara response [see Eq.(\ref{eq:shortwavelengthQSTLS}) of Appendix B] over the short-wavelength limit of the normalized ideal Matsubara density response [see Eq.(\ref{eq:shortwavelengthLindhard}) of Appendix A], directly yields the short-wavelength limit of the qSTLS dynamic Matsubara LFC
\begin{align}
\displaystyle\lim_{x\to\infty}\widetilde{G}_{\mathrm{qSTLS}}(x,l)=1-g(0)\,,\label{eq:shortwavelengthQSTLSLFC}
\end{align}
where $g(0)=1+(3/2)\int_0^{\infty}dyy^2[S(y)-1]$ is the on-top value of the pair correlation function of the qSTLS scheme. Again, it is not surprising that the short-wavelength limit can be expressed through the contact value of the pair correlation function. It is pointed out that also the short-wavelength limit of the qSTLS dynamic Matsubara LFC does not depend on the order of the Matsubara frequency. This result was originally derived by Holas \& Rahman for the ground state UEG\cite{HolasPRB1987}.

\textbf{\emph{The high Matsubara frequency limit}}. The ratio of the high Matsubara frequency limit of the normalized qSTLS auxiliary Matsubara response [see Eq.(\ref{eq:highMatsubaraQSTLS}) of Appendix B] over the high Matsubara frequency limit of the normalized ideal Matsubara density response [see Eq.(\ref{eq:highMatsubaraLindhard}) of Appendix A], leads to the high Matsubara frequency limit of the qSTLS dynamic Matsubara LFC
\begin{align}
\displaystyle\lim_{l\to\infty}\widetilde{G}_{\mathrm{qSTLS}}(x,l)={G}_{\mathrm{STLS}}(x)\,,\label{eq:highMatsubaraQSTLSLFC}
\end{align}
where ${G}_{\mathrm{STLS}}(x)$ is the STLS static LFC functional. It is emphasized that this is not the static LFC of the STLS scheme, because the involved SSF stems from the qSTLS scheme and not from the STLS scheme. This result was also originally derived by Holas \& Rahman for the ground state UEG and in the real frequency domain\cite{HolasPRB1987}. 

\textbf{\emph{The zero Matsubara frequency limit}}. By dividing the static limit of the normalized qSTLS auxiliary Matsubara response [see Eq.(\ref{eq:zeroMatsubaraQSTLS}) of Appendix B] with the static limit of the normalized ideal Matsubara density response [see Eq.(\ref{eq:zeroMatsubaraLindhard}) of Appendix A], we obtain the zero Matsubara frequency limit of the qSTLS dynamic Matsubara LFC
\begin{align}
\displaystyle\lim_{l\to0}\widetilde{G}_{\mathrm{qSTLS}}(x,l)=-\frac{3}{4}\frac{\displaystyle\int_0^{\infty}w\left[S(w)-1\right]dw\int_0^{\infty}dy\frac{y\exp{\left(\frac{y^2}{\Theta}-\bar{\mu}\right)}}{\left[\exp{\left(\frac{y^2}{\Theta}-\bar{\mu}\right)}+1\right]^2}\int_{x^2-xw}^{x^2+xw}\frac{xdt}{2t+w^2-x^2}\left[\left(y^2-\frac{t^2}{4x^2}\right)\ln{\left|\frac{2xy+t}{2xy-t}\right|}+\frac{y}{x}t\right]}{\displaystyle\int_0^{\infty}dy\frac{y\exp{\left(\frac{y^2}{\Theta}-\bar{\mu}\right)}}{\left[\exp{\left(\frac{y^2}{\Theta}-\bar{\mu}\right)}+1\right]^2}\left[\left(y^2-\frac{x^2}{4}\right)\ln{\left|\frac{x+2y}{x-2y}\right|}+xy\right]}\,.\label{eq:zeroMatsubaraQSTLSLFC}
\end{align}
This result was originally derived by Schweng \& B\"ohm\cite{SchwengPRB1993}.

\textbf{\emph{The combined long-wavelength and high Matsubara frequency limit}}. The sequential limit $l\to\infty,x\to0$ is trivial courtesy of Eq.(\ref{eq:longwavelengthQSTLSLFC}). The sequential limit $x\to0,l\to\infty$ is computed from Eq.(\ref{eq:highMatsubaraQSTLSLFC}) and the known long-wavelength limit of the STLS functional ${G}_{\mathrm{STLS}}(x\to0)=-(1/2)\pi\lambda{r}_{\mathrm{s}}\widetilde{u}_{\mathrm{int}}x^2$\cite{TanakaJPSJ1986}. The sequential limits are equal and, thus, this double limit of the qSTLS dynamic Matsubara LFC exists and is given by
\begin{align}
\displaystyle\lim_{x\to0}\displaystyle\lim_{l\to\infty}\widetilde{G}_{\mathrm{qSTLS}}(x,l)=\displaystyle\lim_{l\to\infty}\displaystyle\lim_{x\to0}\widetilde{G}_{\mathrm{qSTLS}}(x,l)=-\frac{1}{2}\pi\lambda{r}_{\mathrm{s}}\widetilde{u}_{\mathrm{int}}x^2\,.\label{eq:longwavelengthhighMatsubaraQSTLSLFC}
\end{align}

\textbf{\emph{The combined short-wavelength and high Matsubara frequency limit}}. The sequential limit $l\to\infty,x\to\infty$ is trivial courtesy of Eq.(\ref{eq:shortwavelengthQSTLSLFC}). The sequential limit $x\to\infty,l\to\infty$ is computed from Eq.(\ref{eq:highMatsubaraQSTLSLFC}) and the known short-wavelength limit of the STLS functional ${G}_{\mathrm{STLS}}(x\to\infty)=1-g(0)$\cite{TanakaJPSJ1986}. The sequential limits are equal and, thus, this double limit of the qSTLS dynamic Matsubara LFC also exists and is given by
\begin{align}
\displaystyle\lim_{x\to\infty}\displaystyle\lim_{l\to\infty}\widetilde{G}_{\mathrm{qSTLS}}(x,l)=\displaystyle\lim_{l\to\infty}\displaystyle\lim_{x\to\infty}\widetilde{G}_{\mathrm{qSTLS}}(x,l)=1-g(0)\,.\label{eq:shortwavelengthhighMatsubaraQSTLSLFC}
\end{align}

\textbf{\emph{The combined short-wavelength and zero Matsubara frequency limit}}. The sequential limit $l\to0,x\to\infty$ is apparently equal to the single limit $x\to\infty$, courtesy of Eq.(\ref{eq:shortwavelengthQSTLSLFC}). It is also straightforward to deduce that the sequential limit $x\to\infty,l\to0$ is also equal to the single limit $x\to\infty$. The sequential limits are equal and, thus, this double limit of the qSTLS dynamic Matsubara LFC also exists and is given by
\begin{align}
\displaystyle\lim_{x\to\infty}\displaystyle\lim_{l\to0}\widetilde{G}_{\mathrm{qSTLS}}(x,l)=\displaystyle\lim_{l\to0}\displaystyle\lim_{x\to\infty}\widetilde{G}_{\mathrm{qSTLS}}(x,l)=1-g(0)\,.\label{eq:shortwavelengthzeroMatsubaraQSTLSLFC}
\end{align}

\textbf{\emph{The combined long-wavelength and zero Matsubara frequency limit}}. The sequential limit $l\to0,x\to0$ is trivial courtesy of Eq.(\ref{eq:longwavelengthQSTLSLFC}). It reads as
\begin{align}
\displaystyle\lim_{l\to0}\lim_{x\to0}\widetilde{G}_{\mathrm{qSTLS}}(x,l)=-\frac{1}{2}\pi\lambda{r}_{\mathrm{s}}\widetilde{u}_{\mathrm{int}}x^2\,,\label{eq:zeroMatsubaralongwavelengthQSTLSLFC}
\end{align}
The sequential limit $x\to0,l\to0$ is computed by dividing Eq.(\ref{eq:longwavelengthzeroMatsubaraQSTLS}) of Appendix B with Eq.(\ref{eq:longwavelengthzeroMatsubaraLindhard}) of Appendix A. This leads to
\begin{align}
\displaystyle\lim_{x\to0}\lim_{l\to0}\widetilde{G}_{\mathrm{qSTLS}}(x,l)=\frac{\displaystyle6x^2\int_0^{\infty}\left[S(w)-1\right]dw\int_0^{\infty}\frac{dy}{\exp{\left(\frac{y^2}{\Theta}-\bar{\mu}\right)}+1}\left[\frac{y^2}{w^2}-\frac{y^3}{w^3}\ln{\left|\frac{w+2y}{w-2y}\right|}\right]}{\displaystyle\int_0^{\infty}dy\frac{1}{\exp{\left(\frac{y^2}{\Theta}-\bar{\mu}\right)}+1}}\,.\label{eq:longwavelengthzeroMatsubaraQSTLSLFC}
\end{align}
Only in this case, the sequential limits differ and, thus, this double limit of the qSTLS dynamic Matsubara LFC does not exist.

\subsection{Limiting behavior of the exact dynamic Matsubara local field correction}\label{sec:exactasymptotics}

It is useful to repeat some expressions for the limiting behavior of the exact dynamic Matsubara LFC. A comparison with the aforementioned expressions for the limiting behavior of the qSTLS dynamic Matsubara LFC suffices to reveal some of the shortcomings of the qSTLS scheme without even requiring its numerical solution.

The exact long wavelength limit of the static LFC of the finite temperature UEG is the most well-known expression that involves the local field correction and is described by the compressibility sum rule\cite{GiulianiVignale,SingwiSSP1981}. It is also trivial to show that $G(\boldsymbol{q},\omega\to0)=\widetilde{G}(\boldsymbol{q},l\to0)$. Thus, in normalized units and after expanding the second-order density derivative, the \emph{exact long wavelength limit of the static Matsubara LFC} reads as\cite{ToliasPRB2024,DornheimPRB2021}
\begin{align}
\displaystyle\lim_{x\to0}\lim_{l\to0}\widetilde{G}(x,l)=-\frac{\pi}{12}\lambda{r}_{\mathrm{s}}\left(4\Theta^2\frac{\partial^2}{\partial{\Theta}^2}+r_{\mathrm{s}}^2\frac{\partial^2}{\partial{r_{\mathrm{s}}^2}}+4\Theta{r}_{\mathrm{s}}\frac{\partial^2}{\partial{\Theta}\partial{r_{\mathrm{s}}}}-2\Theta\frac{\partial}{\partial{\Theta}}-2r_{\mathrm{s}}\frac{\partial}{\partial{r_{\mathrm{s}}}}\right)\widetilde{f}_{\mathrm{xc}}x^2\,.\label{eq:longwavelengthzeroMatsubaraexactLFC}
\end{align}
where $\widetilde{f}_{\mathrm{xc}}$ is the exchange--correlation contribution to the free energy per electron (expressed in Hartree energy units) that can be computed directly from the interaction energy with the aid of the adiabatic--connection formula\cite{GiulianiVignale}.

The exact high frequency limit of the dynamic LFC of the finite temperature UEG has been known for $50$ years\cite{PathakPRB1973,NiklassonPRB1974,KuglerJSP1975,HolasCollection1987}. Its derivation is perhaps the most well known application of the theory of frequency moment sum rules\cite{GiulianiVignale,KuglerJSP1975}. It is based on the high frequency expansion of the constitutive relation that defines the LFC [see Eq.(\ref{eq:densityresponseDLFC})] combined with the f-sum rule and the third frequency moment sum rule\cite{GiulianiVignale,KuglerJSP1975}. It is straightforward to show that the high frequency limit of the dynamic LFC is equal to the high Matsubara frequency limit of the dynamic Matsubara LFC, i.e., $G(\boldsymbol{q},\omega\to\infty)=\widetilde{G}(\boldsymbol{q},l\to\infty)$\cite{DornheimPRB2024}. Thus, in normalized units, the \emph{exact high Matsubara frequency limit of the dynamic Matsubara LFC} reads as\cite{DornheimPRB2024}
\begin{align}
\displaystyle\lim_{l\to\infty}\widetilde{G}(x,l)= I(x)-\frac{3}{2}\pi\lambda{r}_{\mathrm{s}}\widetilde{T}_{\mathrm{xc}}x^2\,,\label{eq:highMatsubaraexactLFC}
\end{align}
where $\widetilde{T}_{\mathrm{xc}}=\widetilde{T}-\widetilde{T}_{\mathrm{HF}}$ is the exchange--correlation contribution to the kinetic energy per electron (expressed in Hartree energy units) and where $I(x)$ is a functional of the SSF, often referred to as the Pathak-Vashishta functional, that is defined by\cite{SingwiSSP1981,PathakPRB1973,KuglerJSP1975}
\begin{align}
I(x)=-\frac{3}{8}\int_0^{\infty}dyy^2[S(y)-1]\left[\frac{5}{3}-\frac{y^2}{x^2}+\frac{(y^2-x^2)^2}{2yx^3}\ln{\left|\frac{y+x}{y-x}\right|}\right]\,.\label{eq:highMatsubaraexactLFCfun1}
\end{align}
The long and short wavelength limits of the Pathak-Vashishta functional are easily found to be\cite{KuglerJSP1975}
\begin{align}
\displaystyle\lim_{x\to0}I(x)&=-\frac{1}{5}\pi\lambda{r}_{\mathrm{s}}\widetilde{u}_{\mathrm{int}}x^2\,,\label{eq:highMatsubaraexactLFCfun2}\\
\displaystyle\lim_{x\to\infty}I(x)&=\frac{2}{3}\left[1-g(0)\right]\,.\label{eq:highMatsubaraexactLFCfun3}
\end{align}
Combining the above, the \emph{exact long and short wavelength limits of the high Matsubara frequency limit of the dynamic Matsubara LFC} read as\cite{KuglerJSP1975}
\begin{align}
\displaystyle\lim_{x\to0}\lim_{l\to\infty}\widetilde{G}(x,l)&=-\frac{1}{10}\pi\lambda{r}_{\mathrm{s}}\left(2\widetilde{u}_{\mathrm{int}}+15\widetilde{T}_{\mathrm{xc}}\right)x^2\,,\label{eq:longwavelengthhighMatsubaraexactLFC}\\
\displaystyle\lim_{x\to\infty}\lim_{l\to\infty}\widetilde{G}(x,l)&=-\frac{3}{2}\pi\lambda{r}_{\mathrm{s}}\widetilde{T}_{\mathrm{xc}}x^2\,.\label{eq:shortwavelengthhighMatsubaraexactLFC}
\end{align}
It is worth noting that the correctness of Eqs.(\ref{eq:highMatsubaraexactLFC},\ref{eq:longwavelengthhighMatsubaraexactLFC},\ref{eq:shortwavelengthhighMatsubaraexactLFC}) has also been confirmed by PIMC simulations\cite{DornheimPRB2024,DornheimEPL2024}. It is also emphasized that the short wavelength limit of the high Matsubara frequency limit of the dynamic Matsubara LFC diverges for all UEG state points, except for those that belong to the phase diagram line that is defined by $\widetilde{T}_{\mathrm{xc}}(r_{\mathrm{s}},\Theta)=0$.

The exact short wavelength limit of the static LFC of the ground state UEG ($\Theta=0$) has also been known for $50$ years and has been derived from different formalisms\cite{NiklassonPRB1974,HolasCollection1987,VignalePRB1988}. Very recently, the result was generalized to arbitrary degeneracy and even to arbitrary frequencies by utilizing thermodynamic Green's functions within the Matsubara formalism\cite{HouPRB2022}. Thus, the \emph{exact short wavelength limit of the dynamic Matsubara LFC} reads as\cite{HouPRB2022}
\begin{align}
\displaystyle\lim_{x\to\infty}\widetilde{G}(x,l)=\frac{1}{2}\pi\lambda{r}_{\mathrm{s}}\widetilde{T}_{\mathrm{xc}}x^2\,,\label{eq:shortwavelengthexactLFC}
\end{align}
It is worth noting that the correctness of Eq.(\ref{eq:shortwavelengthexactLFC}) has also been confirmed by PIMC simulations\cite{dornheim2024shortwavelengthlimitdynamic}. It is evident that the short wavelength limit of the dynamic Matsubara LFC does not depend on the Matsubara order. It is also emphasized that the short wavelength limit of the dynamic Matsubara LFC diverges for all UEG state points, except for those which satisfy $\widetilde{T}_{\mathrm{xc}}(r_{\mathrm{s}},\Theta)=0$. Finally, a comparison of Eqs(\ref{eq:shortwavelengthhighMatsubaraexactLFC},\ref{eq:shortwavelengthexactLFC}) directly reveals that the double short wavelength limit and high Matsubara frequency limit of the dynamic Matsubara LFC does not formally exist.

\begin{figure}[b]
	\centering
	\includegraphics[width=7.20in]{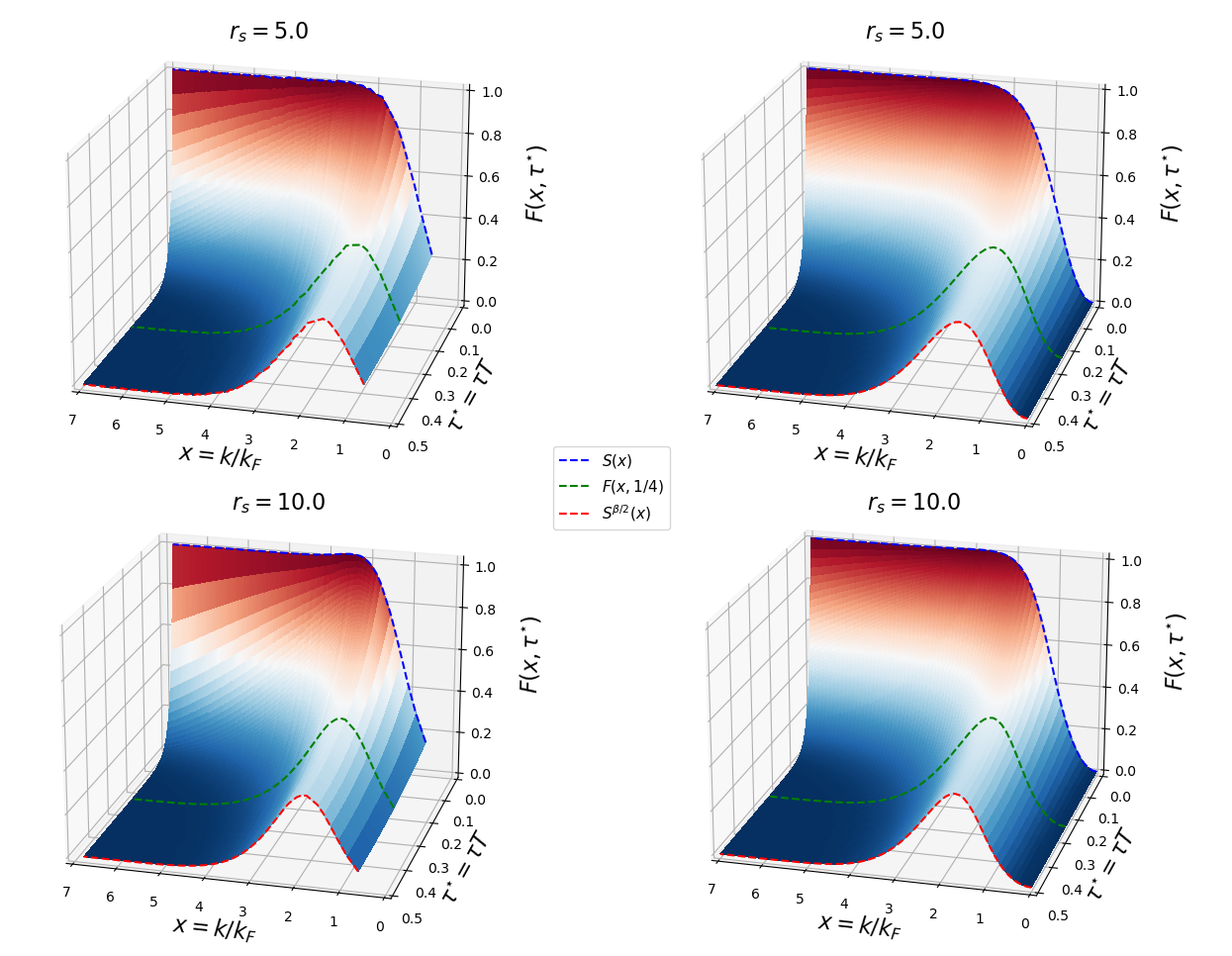}
	\caption{The imaginary--time correlation function of the paramagnetic uniform electron gas at $\Theta=1$ and $r_{\mathrm{s}}=5$ (first row) and $r_{\mathrm{s}}=10$ (second row). Shown are results from PIMC simulations (left panel) and the qSTLS scheme (right panel). The static structure factor, the imaginary--time correlation function when $\tau^{\star}=1/4$ and the thermal structure factor are demarcated with blue, green and red dashed lines, respectively.}\label{fig:ITCFfull}
\end{figure}

\section{Numerical results}

\subsection{Imaginary-time domain responses}

As aforementioned, in PIMC simulations, the ITCFs are directly accessible in the interval $\tau\in[0,\beta]$ or $\tau^{\star}\in[0,1]$
from their definition $F(\boldsymbol{k},\tau)=(1/N)\left\langle\hat{\rho}_{\boldsymbol{k}}(-\imath\hbar\tau)\hat{\rho}^{\dagger}_{\boldsymbol{k}}\right\rangle_0$. On the other hand, in the dielectric formalism, the ITCFs are computed from the Fourier--Matsubara expansion with the converged SSFs [see Eq.(\ref{qSTLSpost3})]. It is pointed out that the detailed balance condition leads to the imaginary--time symmetry property ${F}(\boldsymbol{k},\tau+\beta/2)=F(\boldsymbol{k},\beta/2-\tau)$ for the ITCF\cite{DornheimPoPRev2023,DornheimMRE2023,DornheimNat2022}. This property directly yields that the ITCF has a minimum at $\tau=\beta/2$ regardless of the wavenumber. This property also implies that it suffices to study the ITCF in the imaginary--time internal $\tau\in[0,\beta/2]$ or $\tau^{\star}\in[0,1/2]$, as done herein. Results from PIMC simulations and the qSTLS scheme for the imaginary--time wavenumber resolved ITCFs are illustrated in Fig.\ref{fig:ITCFfull}, for $\Theta=1$ and $r_{\mathrm{s}}=5,\,10$. 

This figure does not facilitate a graphical comparison and has been added for completeness. In what follows, we shall confine ourselves to the imaginary--time slices $\tau^{\star}=0,1/4,1/2$. Nevertheless, our conclusions shall be valid for any imaginary--time. The quasi-exact PIMC results will be compared to the results of first-principles dielectric schemes (RPA, STLS, qSTLS) and to the results of the effective static approximation (ESA)\cite{DornheimPRL2020,DornheimPRB2021}. The ESA scheme constructs a closed-form
semi-empirical expression for a (frequency-averaged) static LFC based on the exact long and short wavelength limits as well as on a neural net representation of quantum Monte Carlo results at intermediate wavenumbers\cite{DornheimJCP2019}. The ESA scheme is expected to be highly accurate within the WDM regime. Its inclusion aims to demonstrate that, despite its static LFC, the semi-empirical ESA scheme leads to highly accurate predictions for the dynamic properties of the UEG. Owing to the sparsity of PIMC results for the DSF (due to the AC problem)\cite{DornheimPRL2018,GrothPRB2019,DornheimPRB2020b}, the dynamic aspect of ESA has been poorly investigated in the past\cite{DornheimPRB2021}. The Fourier--Matsubara expansion of the ITCF makes such an investigation a rather trivial matter.

\begin{figure}
	\centering
	\includegraphics[width=7.00in]{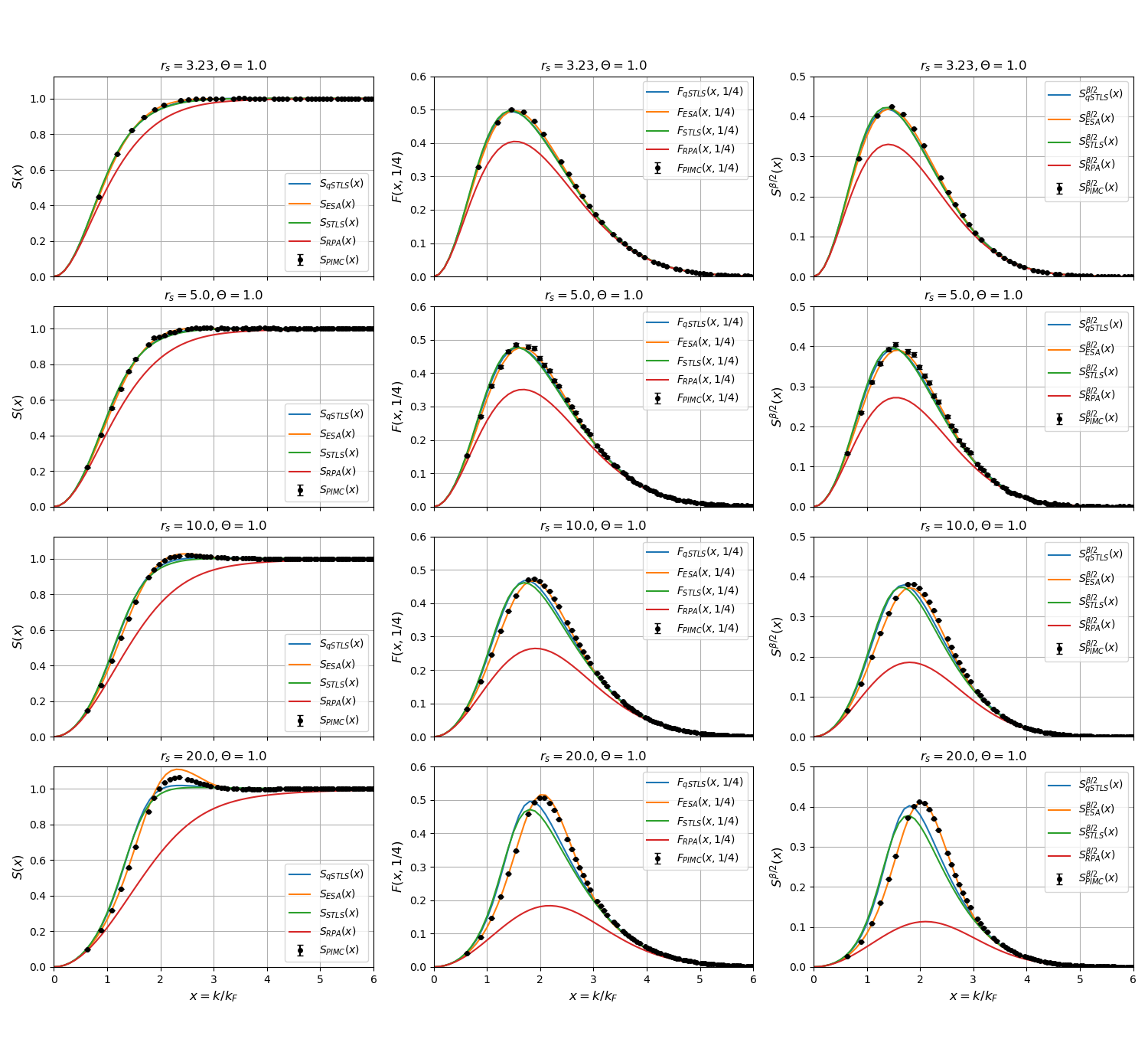}
	\caption{The static structure factor (left panel), the imaginary--time correlation function when $\tau^{\star}=1/4$ (middle panel) and the thermal structure factor (right panel) of the paramagnetic uniform electron gas at $\Theta=1$ and $r_{\mathrm{s}}=3.23$ (first row), $r_{\mathrm{s}}=5$ (second row), $r_{\mathrm{s}}=10$ (third row), $r_{\mathrm{s}}=20$ (fourth row). Shown are results from PIMC simulations (black symbols), the RPA scheme (red solid lines), the STLS scheme (green solid lines), the ESA scheme (orange solid lines) and the qSTLS scheme (blue solid lines).}\label{fig:ITCFcuts}
\end{figure}

A comparison of the quasi-exact SSF from PIMC simulations with the SSF predictions of the RPA, STLS, qSTLS and ESA schemes is featured in the left panel of Fig.\ref{fig:ITCFcuts}. The UEG state points are all at the Fermi temperature with the coupling parameter gradually increasing from $r_{\mathrm{s}}=3.23$ up to $r_{\mathrm{s}}=20$. \textbf{(i)} As expected, the ESA-generated SSF is nearly indistinguishable from the PIMC SSF for all $4$ coupling parameters. However, for $r_{\mathrm{s}}=20$, there is a clear ESA overestimation of the PIMC result that is confined to the vicinity of the SSF maximum\cite{DornheimJCP2021}. \textbf{(ii)} As expected, the RPA-generated SSF is strongly underestimating the PIMC SSF, with the deviations growing with the coupling parameter. Therefore, even for $r_{\mathrm{s}}=3.23$, local field corrections are rather essential for the correct description of the static structure. \textbf{(iii)} The STLS- and qSTLS-generated SSFs are essentially overlapping for $r_{\mathrm{s}}=3.23,\,5$. At these state points, they are nearly indistinguishable from the PIMC and ESA results. However, for $r_{\mathrm{s}}=10,\,20$, the qSTLS-generated SSFs become visibly more accurate, but they still underestimate the PIMC result in the vicinity of the maximum. \textbf{(iv)} Although barely visible in the graphs, for $r_{\mathrm{s}}=20$, both the qSTLS and STLS SSFs exceed unity in the intermediate wavenumber range and, thus, attain a rather broad maximum prior to reaching their asymptotic limit of unity.

A comparison of the quasi-exact TSF from PIMC simulations with the TSF predictions of the RPA, STLS, qSTLS and ESA schemes is featured in the right panel of Fig.\ref{fig:ITCFcuts} for the same state points. \textbf{(i)} The ESA-generated TSF is nearly indistinguishable from the PIMC TSF for all $4$ coupling parameters. In contrast to the SSF, for $r_{\mathrm{s}}=5,\,10$, there is a small ESA underestimation of the PIMC result that is confined to the vicinity of the TSF maximum. \textbf{(ii)} The RPA-generated TSF is strongly underestimating the PIMC TSF, with the deviations growing with the coupling parameter. The level of underestimation of the TSF is much larger than the level of underestimation of the SSF. Thus, local field effects appear to be more impactful for dynamic density-density correlations than equal time ones. \textbf{(iii)} The STLS- and qSTLS-generated TSFs are nearly overlapping and are very close to the PIMC and ESA results for $r_{\mathrm{s}}=3.23,\,5$. However, for $r_{\mathrm{s}}=10,\,20$, the qSTLS-generated TSFs become visibly more accurate, but their deviations from the PIMC TSFs are notable and spread over the intermediate wavenumber range of $k_{\mathrm{F}}\lesssim{k}\lesssim3{k}_{\mathrm{F}}$. \textbf{(iv)} It is worth mentioning that the qSTLS scheme consistently improves the STLS scheme prediction for the magnitude of the TSF maximum, but barely affects the STLS prediction for the position of the TSF maximum. \textbf{(v)} A comparison in terms of the ITCF evaluated at $\tau^{\star}=1/4$ is featured in the middle panel of Fig.\ref{fig:ITCFcuts}. It is evident that the TSF discussion also applies to $F(x,1/4)$ and to the entire ITCF $F(x,\tau^{\star})$.

The substitution of the classical BBGKY hierarchy (STLS scheme) with the quantum BBGKY hierarchy (qSTLS scheme) would be naively expected to have a strong impact on the structural and dynamic predictions of the dielectric formalism for warm dense UEG states of moderate degeneracy. However, at the Fermi temperature, it is evident that the qSTLS scheme only offers a marginal improvement over the STLS scheme. In fact, the phase diagram region where the improvement is meaningful belongs to the boundary between the strongly coupled regime and the WDM regime. For completeness, we note that it can be expected that the qSTLS scheme leads to more substantial improvements to the STLS scheme at high degeneracy, $\Theta\lesssim0.5$. However, such state points will not be analyzed due to the lack of PIMC benchmark data. 

\subsection{Matsubara frequency space responses}

\begin{figure}
	\centering
	\includegraphics[width=7.00in]{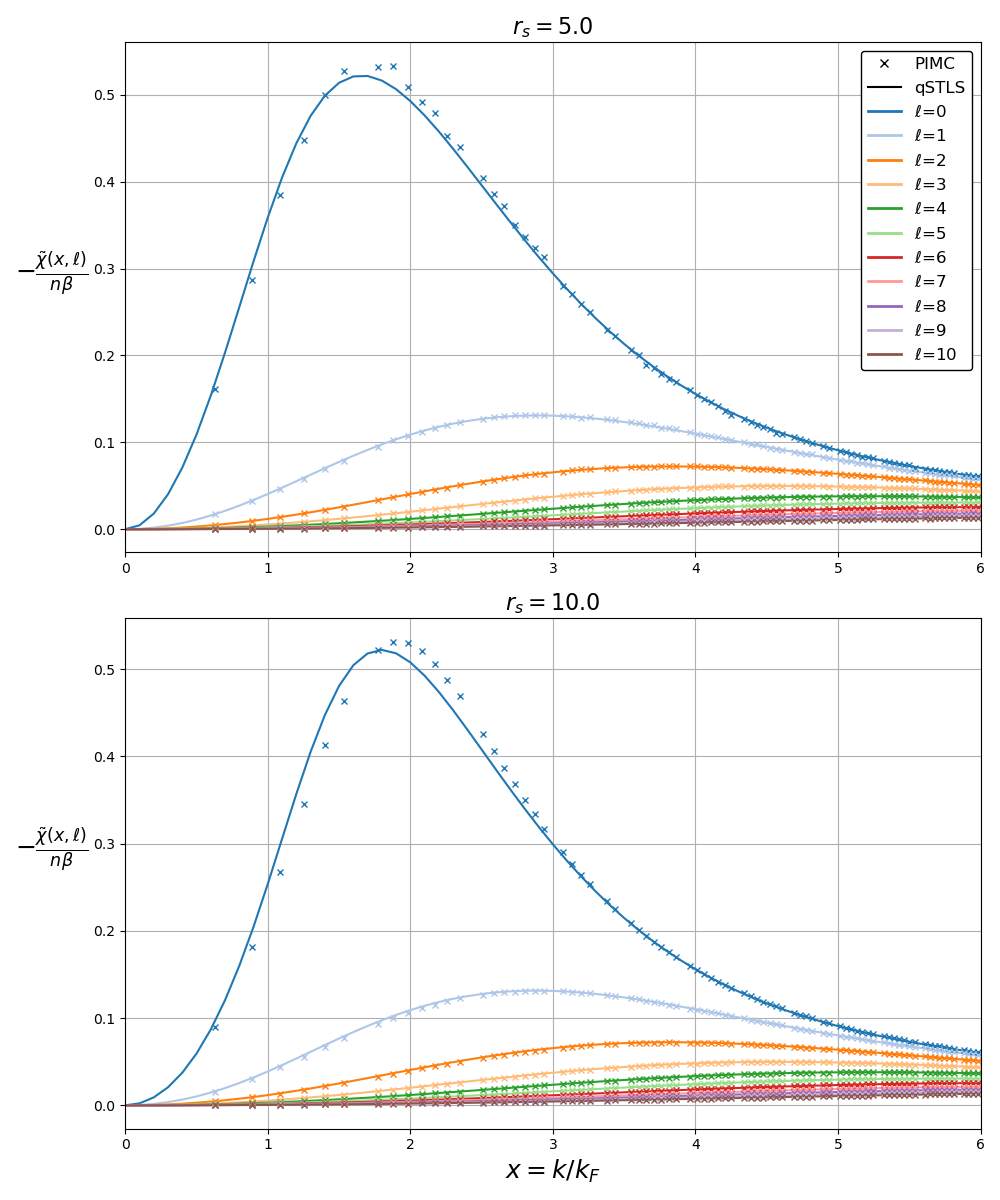}
	\caption{The dynamic Matsubara density response of the paramagnetic uniform electron gas at $\Theta=1$ and $r_{\mathrm{s}}=5$ (top), $r_{\mathrm{s}}=10$ (bottom) as a function of the wavenumber. Shown are results for different Matsubara orders $l=0,1,\cdots,10$  from the qSTLS scheme (solid lines) and from PIMC simulations (symbols).}\label{fig:MAtsubaraDSF}
\end{figure}

As aforementioned, in PIMC simulations, the dynamic Matsubara density responses of arbitrary order are obtained from the inverted Fourier--Matsubara expansion with knowledge of the ITCF [see Eq.(\ref{eq:invDRFFourierMatsubaraSeries})]. On the other hand, in the dielectric formalism, the dynamic Matsubara density responses of arbitrary order are automatically obtained from the computational loop with the converged SSF. Results from PIMC simulations and the qSTLS scheme for the wavenumber resolved dynamic Matsubara density response at different orders of the Matsubara frequency ($l=0-10$) are illustrated in Fig.\ref{fig:MAtsubaraDSF}, for $\Theta=1$ and $r_{\mathrm{s}}=5,\,10$. Owing to the dramatic change in magnitude between the dynamic Matsubara density responses evaluated at successive orders $l$, this plot essentially only allows a comparison for the static density response. \textbf{(i)} Unsurprisingly, the qSTLS-generated static density responses are very accurate in the short wavelength limit (which is dictated by the ideal static density response) and the long wavelength limit (which is dictated by the perfect screening). \textbf{(ii)} Deviations between the qSTLS- and PIMC-generated static density responses emerge at the vicinity of the minimum. For $r_{\mathrm{s}}=5$, the qSTLS scheme leads to a slight overestimation of the absolute value for $k\sim{k}_{\mathrm{F}}$ (wavenumbers smaller than the minimum position) and to a slight underestimation of the absolute value for $k\sim2{k}_{\mathrm{F}}$  (wavenumbers larger than and around the minimum position). For $r_{\mathrm{s}}=10$, these deviations grow and it also becomes evident that there is an offset at the predicted position of the minimum. \textbf{(iii)} In terms of higher order dynamic Matsubara density responses, these plots only reveal that the qSTLS-generated curves generally follow their PIMC-generated counterparts without large deviations. 

In order to better appreciate the discrepancies, it is preferable to focus on the dynamic Matsubara LFCs which constitute a measure of the deviations between the dynamic Matsubara density responses and the ideal dynamic Matsubara density responses. As aforementioned, in PIMC simulations, the dynamic Matsubara LFCs of arbitrary order are obtained from the constitutive relation with knowledge of the dynamic Matsubara density response of the same order [see Eq.(\ref{eq:invLFCFourierMatsubaraSeries})]. On the other hand, in the dielectric formalism, the dynamic Matsubara LFCs are again automatically obtained from the computational loop with the converged SSF.

\begin{figure}
	\centering
	\includegraphics[width=7.00in]{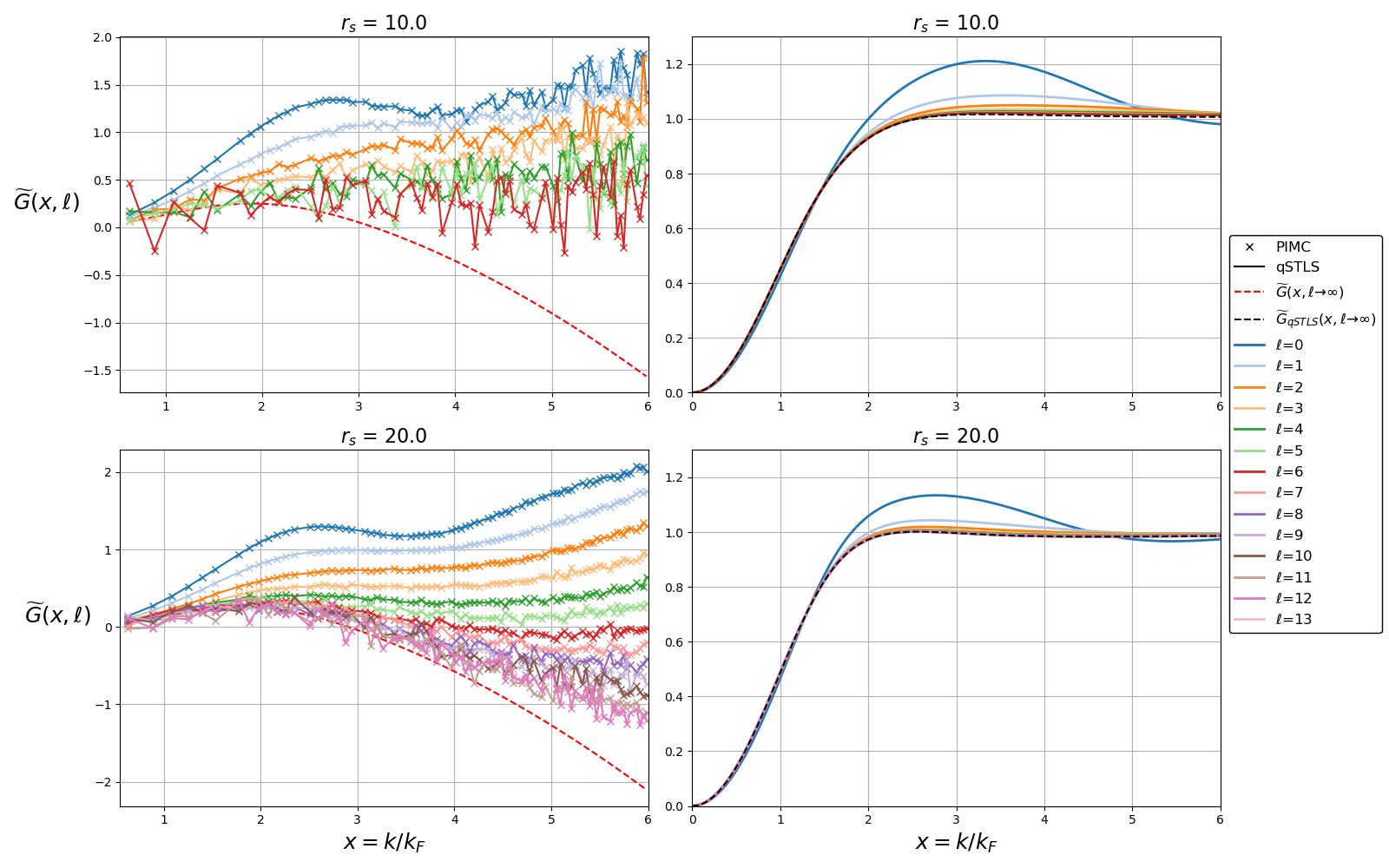}
	\caption{The dynamic Matsubara local field correction of the paramagnetic uniform electron gas at $\Theta=1$ as a function of the wavenumber. Top: Results for $r_{\mathrm{s}}=10$ at different Matsubara orders $l=0,1,\cdots,6$ from PIMC simulations (left) and from the qSTLS scheme (right). Bottom: Results for $r_{\mathrm{s}}=20$ at different Matsubara orders $l=0,1,\cdots,13$ from PIMC simulations (left) and from the qSTLS scheme (right). The exact high Matsubara frequency limit [see Eq.(\ref{eq:highMatsubaraexactLFC})] and the qSTLS high Matsubara frequency limit [see Eq.(\ref{eq:highMatsubaraQSTLSLFC})] are also added.}\label{fig:MAtsubaraLFC}
\end{figure}

Let us first focus on the quasi-exact dynamic Matsubara LFCs from the PIMC simulations, i.e., the left panel of Fig.\ref{fig:MAtsubaraLFC}. \textbf{(i)} The exact high Matsubara frequency limit of the dynamic Matsubara LFC is computed from Eq.(\ref{eq:highMatsubaraexactLFC}). The Pathak-Vashishta functional $I(x)$ is evaluated with the exact PIMC SSF, which is properly extrapolated towards zero at the inaccessible long wavelengths. The exchange-correlation contribution to the kinetic energy $\widetilde{T}_{\mathrm{xc}}$ is evaluated from the highly accurate GDSMFB parametrization of the exchange correlation free energy $\widetilde{f}_{\mathrm{xc}}$\cite{DornheimPhysRep2018,GrothPRL2017}, after the necessary thermodynamic differentiations with respect to $r_{\mathrm{s}}$ and $\Theta$\cite{DornheimPhysRep2018}. \textbf{(ii)} With increasing order of the Matsubara frequency $l$, the deviations between the equal-order dynamic Matsubara density response and ideal dynamic Matsubara density response become smaller. Thus, the dynamic Matsubara LFC data become more noisy as $l$ increases. As a consequence, the first $7$ Matsubara orders $l=0,1,\cdots,6$ are well resolved for $\Theta=1,\,r_{\mathrm{s}}=10$ and the first $14$ Matsubara orders $l=0,1,\cdots,13$ are well resolved for $\Theta=1,\,r_{\mathrm{s}}=20$. \textbf{(iii)} The dynamic Matsubara LFCs for finite $l$ are consistent with the short wavelength limit divergence implied by Eq.(\ref{eq:shortwavelengthexactLFC}). Note that $\widetilde{T}_{\mathrm{xc}}>0$ for the investigated UEG state points. More detailed investigations have revealed that the short wavelength limit is approached at higher wavenumbers as the Matsubara order increases\cite{dornheim2024shortwavelengthlimitdynamic}.
\textbf{(iv)} The resolved Matsubara orders suffice to show that, at these state points, the dynamic Matsubara LFC is bounded from above by its static limit $l\to0$ and bounded from below by its high Matsubara frequency limit $l\to\infty$. Owing to the differences between these two limits, the quasi-exact dynamic Matsubara LFC exhibits appreciable variations regardless of the wavenumber.

We proceed with the qSTLS dynamic Matsubara LFCs, i.e., the right panel of Fig.\ref{fig:MAtsubaraLFC}. \textbf{(i)} The high Matsubara frequency limit of the qSTLS dynamic Matsubara LFC is computed from Eq.(\ref{eq:highMatsubaraQSTLSLFC}), which is the STLS LFC functional but evaluated with the qSTLS SSF. Owing to the small differences between the STLS and qSTLS schemes, the high Matsubara frequency limit of the qSTLS dynamic Matsubara LFC is very close to the static limit of the qSTLS dynamic Matsubara LFC. \textbf{(ii)} Regardless of the Matsubara frequency order, the qSTLS dynamic Matsubara LFCs have a common long wavelength limit [see Eq.(\ref{eq:longwavelengthQSTLSLFC})], which extends up to $k\simeq1.5k_{\mathrm{F}}$ for the investigated UEG state points.  \textbf{(iii)} Regardless of the Matsubara frequency order, the qSTLS dynamic Matsubara LFCs have a common short wavelength limit [see Eq.(\ref{eq:shortwavelengthQSTLSLFC})], which is close to unity and begins from $k\simeq5k_{\mathrm{F}}$ for the investigated UEG state points. \textbf{(iv)} Ignoring the weak oscillations of the static limit, it can be stated that the qSTLS dynamic Matsubara LFC is bounded from above by its static limit $l\to0$ and bounded from below by its high Matsubara frequency limit $l\to\infty$. \textbf{(v)} Regardless of the wavenumber, as the Matsubara order increases, there is a rapid convergence from the static limit $l\to0$ to the high Matsubara frequency limit $l\to\infty$.

Overall, it is evident that the qSTLS scheme only manages to adequately capture the $l=0$ and $l=1$ dynamic Matsubara LFCs away from the short wavelength range. This seems to suffice for an adequate treatment of both static and dynamic density--density correlations away from short distances.

\section{Summary and discussion}

The qSTLS scheme, being the archetypal dynamic dielectric scheme, has been utilized for an investigation of the dynamic properties of the paramagnetic warm dense uniform electron gas. In particular, the qSTLS dynamic properties are analyzed in the imaginary--time domain as well as the Matsubara frequency space, benefitting from an exact Fourier--Matsubara expansion of imaginary--time correlation functions\cite{ToliasJCP2024}, and are compared with quasi-exact PIMC benchmark data\cite{DornheimPRB2024,DornheimEPL2024,dornheim2024shortwavelengthlimitdynamic}. Furthermore, the asymptotic limits of the qSTLS dynamic Matsubara local field correction are derived and compared to the exact asymptotic limits of the dynamic Matsubara local field correction. Based on the numerical and theoretical results, the following physics questions can be formulated and partially addressed.

\emph{Why does the qSTLS scheme only marginally improve the STLS scheme?} Despite the utilization of the same closure condition, due to the substitution of the classical BBGKY hierarchy with the quantum BBGKY hierarchy, the qSTLS scheme constitutes a substantial theoretical upgrade over the STLS scheme. This also comes at a considerable computational cost, since the qSTLS functional can be reduced to a triple integral while the STLS functional can be reduced to a single integral (for the UEG and for any isotropic potential) and since the dynamic qSTLS functional requires a re-evaluation at any Matsubara order $l$ while the static STLS functional does not depend on the Matsubara order (with $l=500$ Matsubara orders required for a stringent convergence at all wavenumbers). Nevertheless, comparison with quasi-exact PIMC results directly in the imaginary--time domain, revealed that the qSTLS scheme leads to marginal improvements in the structural and even the dynamic predictions of the STLS scheme for the warm dense UEG. The analysis of the asymptotic limits of the qSTLS dynamic LFC in the Matsubara frequency space offers a straightforward explanation for this observation. The high Matsubara frequency limit of the qSTLS dynamic Matsubara LFC is given by the static STLS LFC functional, while both the long wavelength and short wavelength limits of the qSTLS dynamic Matsubara LFC are independent of the Matsubara order. In addition, numerical results revealed a rapid convergence of the qSTLS dynamic Matsubara LFC to its high Matsubara frequency limit in the intermediate wavenumber region. As a consequence, it can be concluded that, in practice, the qSTLS dynamic LFC functional behaves as a static LFC functional which is close to the static STLS LFC functional. This conclusion does not apply to the strong coupling regime of the finite temperature UEG, $r_{\mathrm{s}}\gtrsim20$, where quantum versions of semi-classical schemes have been observed to lead to significant improvements\cite{ToliasJCP2023}. 

\emph{How can we reconcile the exact asymptotic limits of the LFC with the asymptotic limits of different schemes of the dielectric formalism?} The exact high frequency limit of the dynamic LFC [see Eq.(\ref{eq:highMatsubaraexactLFC})] and the exact short wavelength limit of the dynamic LFC [see Eq.(\ref{eq:shortwavelengthexactLFC})] involve the exchange-correlation contribution to the kinetic energy $\widetilde{T}_{\mathrm{xc}}$, which is responsible for the divergence as $x\to\infty$. Given the three standard ingredients of the dielectric formalism, it is rather impossible to construct a dielectric scheme whose dynamic LFC has a high frequency limit and a  short wavelength limit that both feature $\widetilde{T}_{\mathrm{xc}}$. The culprit is the constitutive relation of the density response function [see Eq.(\ref{eq:densityresponseDLFC})], which establishes a connection to the ideal (Lindhard) density response at the non-interacting occupation numbers. This constitutive relation essentially defines the dynamic LFC, but it is not consistent with the derivation of the qSTLS density response function. In particular, after the substitution of the STLS ansatz in the quantum BBGKY hierarchy, the linearization is carried out around the equilibrium Wigner distribution function of the interacting system $f_0(\boldsymbol{p})$ and not around the Fermi-Dirac distribution function of the non-interacting system $f_{\mathrm{FD}}(\boldsymbol{p})$. Therefore, the ideal (Lindhard) density response and the three-argument ideal density response should have been evaluated at the exact occupation numbers and thus the substitution of $f_0(\boldsymbol{p})\simeq{f}_{\mathrm{FD}}(\boldsymbol{p})$ constitutes an additional approximation. It is straightforward to deduce that this hidden approximation is responsible for the incorrect asymptotic limits of the LFC, since $\langle\hbar^2k^2/(2m)\rangle_{\mathrm{I}}-\langle\hbar^2k^2/(2m)\rangle_0=T_{\mathrm{xc}}$ where the subscript \enquote{I} stands for averaging over the exact occupation numbers and the subscript \enquote{0} stands for averaging over the non-interacting occupation numbers. Naturally, the reformulation of the constitutive relation implies the re-definition of the dynamic LFC. It is this re-definition that directly implies that the problematic $\widetilde{T}_{\mathrm{xc}}$ terms vanish from the exact high frequency and short wavelength limits of the dynamic LFC, as established in the works of Niklasson\cite{NiklassonPRB1974}, Holas\cite{HolasCollection1987} and Vignale\cite{VignalePRB1988}. In fact, Vignale concludes that the thus re-defined dynamic LFCs provide a better starting point for approximate treatments\cite{VignalePRB1988}. The main drawback is that the equilibrium Wigner distribution function of the interacting system $f_0(\boldsymbol{p})$ (in the phase-space formalism) or the momentum distribution function $n(\boldsymbol{k})$ of the interacting system is not known. Therefore, it should either be extracted from quantum Monte Carlo simulations\cite{MilitzerPRL2002,MilitzerHEDP2019,LarkinCPP2018,HungerPRE2021,DornheimPRE2021} or approximated taking into account its exact asymptotic limit\cite{KimballJPA1975,YasuharaPhA1976,HofmannPRB2013}. The emergence of the on-top value of the pair correlation function, $g(0)$, in the exact asymptotic limit adds an extra layer of complexity, in case fully self-consistent schemes are sought. However, there is a chance that the incorporation of the exact asymptotic limit of the momentum distribution function also corrects for the universal pathology of dielectric schemes; the negativity of the pair correlation function near contact\cite{DornheimPhysRep2018,ToliasPRB2024,ToliasJCP2021}. The above ideas will be pursued in dedicated future works.

\section*{Acknowledgments}

This work was partially supported by the Center for Advanced Systems Understanding (CASUS), financed by Germany’s Federal Ministry of Education and Research (BMBF) and the Saxon state government out of the State budget approved by the Saxon State Parliament. Further support is acknowledged for the CASUS Open Project \emph{Guiding dielectric theories with ab initio quantum Monte Carlo simulations: from the strongly coupled electron liquid to warm dense matter}. This work has received funding from the European Research Council (ERC) under the European Union’s Horizon 2022 research and innovation programme (Grant agreement No. 101076233, "PREXTREME"). Views and opinions expressed are however those of the authors only and do not necessarily reflect those of the European Union or the European Research Council Executive Agency. Neither the European Union nor the granting authority can be held responsible for them. Computations were performed on a Bull Cluster at the Center for Information Services and High-Performance Computing (ZIH) at Technische Universit\"at Dresden, at the Norddeutscher Verbund f\"ur Hoch- und H\"ochstleistungsrechnen (HLRN) under grant mvp00024, and on the HoreKa supercomputer funded by the Ministry of Science, Research and the Arts Baden-W\"urttemberg and by the Federal Ministry of Education and Research.

\renewcommand{\theequation}{A\arabic{equation}}
\setcounter{equation}{0}
\section*{APPENDIX A: Limiting behavior of the normalized ideal Matsubara density response}\label{appendix:Lindhard}

\textbf{\emph{General remark}}. The limiting behavior of the normalized ideal Matsubara density response in terms of the wavenumber and the Matsubara order is generally known. For the sake of completeness, the asymptotic analysis will be repeated herein.\\
\textbf{\emph{The long-wavelength limit}}. In the normalized ideal Matsubara density response, Eq.(\ref{eq:LindhardMatsuNorm}), the argument of the logarithm is expanded and the ratio $4x^3y/[x^4+4x^2y^2+\left(2\pi{l}{\Theta}\right)^2]$ is identified as the small parameter. The Taylor series $\ln{|(1+x)/(1-x)|}\simeq{2x}+2x^3/3+\cdots$ is employed and only the first order term is retained.
\begin{align*}
\frac{\widetilde{\chi}_0(x\to0,l)}{n\beta}=-\frac{3}{4}\frac{\Theta}{x}\int_0^{\infty}dy\frac{y}{\exp{\left(\frac{y^2}{\Theta}-\bar{\mu}\right)}+1}\frac{8x^3y}{x^4+4x^2y^2+\left(2\pi{l}{\Theta}\right)^2}\,.\nonumber
\end{align*}
The approximation $x^4+4x^2y^2+\left(2\pi{l}{\Theta}\right)^2\simeq\left(2\pi{l}{\Theta}\right)^2$ is employed in the denominator and the terms are re-arranged. Thus, the long-wavelength limit of the normalized ideal Matsubara density response reads as
\begin{align*}
\frac{\widetilde{\chi}_0(x\to0,l)}{n\beta}=-6\Theta\frac{x^2}{\left(2\pi{l}{\Theta}\right)^2}\int_0^{\infty}dy\frac{y^2}{\exp{\left(\frac{y^2}{\Theta}-\bar{\mu}\right)}+1}\,.\nonumber
\end{align*}
We now recall the normalization of the Fermi-Dirac distribution function $\int_0^{\infty}dy\{y^2/[\exp{\left(y^2/\Theta-\bar{\mu}\right)}+1]\}=1/3$. Finally, the long-wavelength limit of the normalized ideal Matsubara density response can be compactly rewritten as
\begin{align}
\frac{\widetilde{\chi}_0(x\to0,l)}{n\beta}=-2\Theta\frac{x^2}{\left(2\pi{l}{\Theta}\right)^2}\,.\label{eq:longwavelengthLindhard}
\end{align}
\textbf{\emph{The short-wavelength limit}}. In the normalized ideal Matsubara density response, Eq.(\ref{eq:LindhardMatsuNorm}), both $(2\pi{l}{\Theta})^2$ contributions in the argument of the logarithm are negligible. This directly leads to
\begin{align*}
\frac{\widetilde{\chi}_0(x\to\infty,l)}{n\beta}=-\frac{3}{2}\frac{\Theta}{x}\int_0^{\infty}dy\frac{y}{\exp{\left(\frac{y^2}{\Theta}-\bar{\mu}\right)}+1}\ln{\left|\frac{x+2y}{x-2y}\right|}\,.\nonumber
\end{align*}
The ratio $2y/x$ constitutes the small argument of interest. The Taylor series $\ln{|(1+x)/(1-x)|}\simeq{2x}+2x^3/3+\cdots$ is employed and only the first order term is retained. The terms are re-arranged. Thus, the short-wavelength limit of the normalized ideal Matsubara density response reads as
\begin{align*}
\frac{\widetilde{\chi}_0(x\to\infty,l)}{n\beta}=-6\frac{\Theta}{x^2}\int_0^{\infty}dy\frac{y^2}{\exp{\left(\frac{y^2}{\Theta}-\bar{\mu}\right)}+1}\,.\nonumber
\end{align*}
We now recall the normalization of the Fermi-Dirac distribution function $\int_0^{\infty}dy\{y^2/[\exp{\left(y^2/\Theta-\bar{\mu}\right)}+1]\}=1/3$. Finally, the short-wavelength limit of the normalized ideal Matsubara density response can be compactly rewritten as
\begin{align}
\frac{\widetilde{\chi}_0(x\to\infty,l)}{n\beta}=-2\frac{\Theta}{x^2}\,.\label{eq:shortwavelengthLindhard}
\end{align}
\textbf{\emph{The high Matsubara frequency limit}}. In the normalized ideal Matsubara density response, Eq.(\ref{eq:LindhardMatsuNorm}), the small parameters $\left(2xy+t\right)^2/\left(2{\pi}l\Theta\right)^2$ and $\left(2xy-t\right)^2/\left(2{\pi}l\Theta\right)^2$ are easily identified. The Taylor series $\ln{|(1\pm{x})|}\simeq\pm{x}-x^2/2+\cdots$ is employed and only the first order term is retained.
\begin{align*}
\frac{\widetilde{\chi}_0(x,l\to\infty)}{n\beta}=-\frac{3}{4}\frac{\Theta}{x}\int_0^{\infty}dy\frac{y}{\exp{\left(\frac{y^2}{\Theta}-\bar{\mu}\right)}+1}\left\{\frac{\left(x^2+2xy\right)^2}{\left(2\pi{l}{\Theta}\right)^2}-\frac{\left(x^2-2xy\right)^2}{\left(2\pi{l}\Theta\right)^2}\right\}\,.\nonumber
\end{align*}
After some straightforward algebra and some rearrangements, one obtains
\begin{align*}
\frac{\widetilde{\chi}_0(x,l\to\infty)}{n\beta}&=-6\Theta\frac{x^2}{\left(2\pi{l}\Theta\right)^2}\int_0^{\infty}dy\frac{y^2}{\exp{\left(\frac{y^2}{\Theta}-\bar{\mu}\right)}+1}\,.\nonumber
\end{align*}
As per usual, we now recall the normalization of the Fermi-Dirac distribution function $\int_0^{\infty}dy\{y^2/[\exp{\left(y^2/\Theta-\bar{\mu}\right)}+1]\}=1/3$. Finally, the high Matsubara frequency limit of the normalized ideal Matsubara density response can be compactly rewritten as
\begin{align}
\frac{\widetilde{\chi}_0(x,l\to\infty)}{n\beta}&=-2\Theta\frac{x^2}{\left(2\pi{l}\Theta\right)^2}\,.\label{eq:highMatsubaraLindhard}
\end{align}
\textbf{\emph{The zero Matsubara frequency limit}}. This is essentially the static limit. The normalized ideal Matsubara density response, Eq.(\ref{eq:LindhardMatsuNorm}) directly becomes
\begin{align*}
\frac{\widetilde{\chi}_0(x,l\to0)}{n\beta}=-\frac{3}{4}\frac{\Theta}{x}\int_0^{\infty}dy\frac{1}{\exp{\left(\frac{y^2}{\Theta}-\bar{\mu}\right)}+1}\left[2y\ln{\left|\frac{x+2y}{x-2y}\right|}\right]\,.\nonumber
\end{align*}
There is a logarithmic singularity at $y=x/2$ which can be removed by integration by parts,
\begin{align*}
\frac{\widetilde{\chi}_0(x,l\to0)}{n\beta}=-\frac{3}{4}\frac{\Theta}{x}\int_0^{\infty}dy\frac{1}{\exp{\left(\frac{y^2}{\Theta}-\bar{\mu}\right)}+1}\frac{\partial}{\partial{y}}\left[\left(y^2-\frac{x^2}{4}\right)\ln{\left|\frac{x+2y}{x-2y}\right|}+xy\right]\,.\nonumber
\end{align*}
After some algebra, the zero Matsubara frequency limit of the normalized ideal Matsubara density response is finally rewritten as
\begin{align}
\frac{\widetilde{\chi}_0(x,l\to0)}{n\beta}=-\frac{3}{2}\frac{1}{x}\int_0^{\infty}dy\frac{y\exp{\left(\frac{y^2}{\Theta}-\bar{\mu}\right)}}{\left[\exp{\left(\frac{y^2}{\Theta}-\bar{\mu}\right)}+1\right]^2}\left[\left(y^2-\frac{x^2}{4}\right)\ln{\left|\frac{x+2y}{x-2y}\right|}+xy\right]\,.\label{eq:zeroMatsubaraLindhard}
\end{align}
see also Eq.(\ref{eq:LindhardMatsu0Norm}) of the main text.\\
\textbf{\emph{The combined long-wavelength and zero Matsubara frequency limit}}. It is understood that the static limit is considered prior to the long-wavelength limit. Thus, the normalized ideal Matsubara density response, Eq.(\ref{eq:LindhardMatsuNorm}), first becomes
\begin{align*}
\frac{\widetilde{\chi}_0(x\to0,l\to0)}{n\beta}=-\frac{3}{4}\frac{\Theta}{x}\int_0^{\infty}dy\frac{1}{\exp{\left(\frac{y^2}{\Theta}-\bar{\mu}\right)}+1}\left[2y\ln{\left|\frac{x+2y}{x-2y}\right|}\right]\,.\nonumber
\end{align*}
The small parameter is easily identified to be $x/2y$. The Taylor series $\ln{|(1+x)/(1-x)|}\simeq{2x}+2x^3/3+\cdots$ is employed and only the first order term is retained. Thus, the normalized ideal Matsubara density response is finally rewritten as
\begin{align}
\frac{\widetilde{\chi}_0(x\to0,l\to0)}{n\beta}=-\frac{3}{2}\Theta\int_0^{\infty}dy\frac{1}{\exp{\left(\frac{y^2}{\Theta}-\bar{\mu}\right)}+1}\,.\label{eq:longwavelengthzeroMatsubaraLindhard}
\end{align}

\renewcommand{\theequation}{B\arabic{equation}}
\setcounter{equation}{0}
\section*{APPENDIX B: Limiting behavior of the normalized qSTLS auxiliary Matsubara response}\label{appendix:auxiliary}

\textbf{\emph{General remark}}. When computing the limiting behavior of the normalized qSTLS auxiliary Matsubara response, it is convenient to isolate the $x,l$ dependencies which reside exclusively on the definite integral. Therefore, one has
\begin{align}
\frac{\widetilde{I}_{\mathrm{qSTLS}}(x,l)}{n\beta}=\frac{9}{16}\Theta\int_0^{\infty}w\left[S(w)-1\right]dw\int_0^{\infty}\frac{ydy}{\exp{\left(\frac{y^2}{\Theta}-\bar{\mu}\right)}+1}\widetilde{A}_{\mathrm{qSTLS}}(x,y,w,l)\,,\label{eq:auxqSTLSintegralMatsuNorm0}
\end{align}
where the qSTLS auxiliary definite integral is defined by
\begin{align}
\widetilde{A}_{\mathrm{qSTLS}}(x,y,w,l)=\int_{x^2-xw}^{x^2+xw}\frac{dt}{2t+w^2-x^2}\ln{\left[\frac{\left(2xy+t\right)^2+\left(2{\pi}l\Theta\right)^2}{\left(2xy-t\right)^2+\left(2{\pi}l\Theta\right)^2}\right]}\,.\label{eq:auxqSTLSintegralMatsuNorm}
\end{align}
\textbf{\emph{The long-wavelength limit}}. In the qSTLS auxiliary definite integral, Eq.(\ref{eq:auxqSTLSintegralMatsuNorm}), the integration boundaries not only depend on $x$ but also both become zero in the long-wavelength limit. This will considerably complicate the calculation. First, the argument of the logarithm is expanded and the ratio $4xyt/[t^2+4x^2y^2+\left(2\pi{l}{\Theta}\right)^2]$ is identified as the small parameter. The Taylor series $\ln{|(1+x)/(1-x)|}\simeq{2x}+2x^3/3+\cdots$ is employed and only the first order term is retained.
\begin{align*}
\widetilde{A}_{\mathrm{qSTLS}}(x\to0,y,w,l)=\int_{x^2-xw}^{x^2+xw}\frac{dt}{2t+w^2-x^2}\frac{8xyt}{t^2+4x^2y^2+\left(2{\pi}l\Theta\right)^2}\,.\nonumber
\end{align*}
The parameters $a^2=4x^2y^2+\left(2{\pi}l\Theta\right)^2$ and $b=(w^2-x^2)/2$ are introduced for convenience. The integrand is decomposed into simple fractions.
\begin{align*}
\widetilde{A}_{\mathrm{qSTLS}}(x\to0,y,w,l)=4xy\left[\frac{a^2}{a^2+b^2}\int_{x^2-xw}^{x^2+xw}\frac{dt}{t^2+a^2}-\frac{b}{a^2+b^2}\int_{x^2-xw}^{x^2+xw}\frac{dt}{t+b}+\frac{b}{2(a^2+b^2)}\int_{x^2-xw}^{x^2+xw}\frac{2tdt}{t^2+a^2}\right]\,.\nonumber
\end{align*}
The evaluation of the emerging definite integrals is based on the standard indefinite integrals $\int{dt}/{(t^2+a^2)}=(1/a)\arctan{(t/a)}$, $\int{dt}/({t+b})=\ln{|t+b|}$ and $\int{(2tdt)}/{(t^2+a^2)}=\ln{|t^2+a^2|}$. After some algebra, one obtains
\begin{align*}
\widetilde{A}_{\mathrm{qSTLS}}(x\to0,y,w,l)=4xy\left\{\frac{a^2}{a^2+b^2}\frac{1}{a}\left[\arctan{\left(\frac{x^2+xw}{a}\right)}-\arctan{\left(\frac{x^2-xw}{a}\right)}\right]-\frac{2b}{a^2+b^2}\ln{\left|\frac{x+w}{x-w}\right|}\right.\nonumber\\\quad\left.+\frac{b}{2(a^2+b^2)}\ln{\left|\frac{(x^2+xw)^2+a^2}{(x^2-xw)^2+a^2}\right|}\right\}\,.\nonumber
\end{align*}
The calculation is performed up to the third order in $x$. Consequently, the inverse tangent term Taylor expansion is truncated as $\arctan{(x)}\simeq{x}-x^3/3$, the first logarithmic term Taylor expansion is truncated as $\ln{|(1+x)/(1-x)|}\simeq{2x}+2x^3/3$ and the second logarithmic term Taylor expansion is truncated as $\ln{|1+x|}\simeq{x}$. After some cumbersome algebra, these lead to
\begin{align*}
\widetilde{A}_{\mathrm{qSTLS}}(x\to0,y,w,l)=4xy\left\{\frac{2}{a^2+b^2}\frac{x^3}{w}-\frac{1}{a^2+b^2}\frac{2}{3}\frac{x^3w^3}{a^2}-\frac{2b}{a^2+b^2}\frac{2}{3}\frac{x^3}{w^3}+\frac{2b}{a^2+b^2}\frac{x^3w}{a^2}\right\}\,.\nonumber
\end{align*}
The approximation $2b\simeq{w}^2$ is employed for the combined $\propto{x}^3w^3$ terms leading to another $\propto{x}^3/w$ term.
\begin{align*}
\widetilde{A}_{\mathrm{qSTLS}}(x\to0,y,w,l)=\frac{16}{3}\frac{yx^4}{w}\frac{1}{a^2}\,.\nonumber
\end{align*}
The final result is obtained by introducing the approximation $a^2=4x^2y^2+\left(2{\pi}l\Theta\right)^2\simeq\left(2{\pi}l\Theta\right)^2$.
\begin{align*}
\widetilde{A}_{\mathrm{qSTLS}}(x\to0,y,w,l)=\frac{16}{3}\frac{yx^4}{w}\frac{1}{(2\pi{l}\Theta)^2}\,.\nonumber
\end{align*}
Overall, the long-wavelength limit of the normalized qSTLS auxiliary Matsubara response becomes
\begin{align*}
\frac{\widetilde{I}_{\mathrm{qSTLS}}(x\to0,l)}{n\beta}=3\Theta\frac{x^4}{(2\pi{l}\Theta)^2}\int_0^{\infty}\left[S(w)-1\right]dw\int_0^{\infty}dy\frac{y^2}{\exp{\left(\frac{y^2}{\Theta}-\bar{\mu}\right)}+1}\,.\nonumber
\end{align*}
At this point, we recall that the interaction energy is given by $\widetilde{u}_{\mathrm{int}}=[1/(\pi\lambda{r}_{\mathrm{s}})]\int_0^{\infty}\left[S(w)-1\right]dw$ in Hartree energy units. Therefore, the long-wavelength limit of the normalized qSTLS auxiliary Matsubara response becomes
\begin{align*}
\frac{\widetilde{I}_{\mathrm{qSTLS}}(x\to0,l)}{n\beta}=3\Theta\frac{x^4}{(2\pi{l}\Theta)^2}\left(\pi\lambda{r}_{\mathrm{s}}\right)\widetilde{u}_{\mathrm{int}}\int_0^{\infty}dy\frac{y^2}{\exp{\left(\frac{y^2}{\Theta}-\bar{\mu}\right)}+1}\,.\nonumber
\end{align*}
We also recall the normalization of the Fermi-Dirac distribution function $\int_0^{\infty}dy\{y^2/[\exp{\left(y^2/\Theta-\bar{\mu}\right)}+1]\}=1/3$. Thus, finally, the long-wavelength limit of the normalized qSTLS auxiliary Matsubara response can be compactly rewritten as
\begin{align}
\frac{\widetilde{I}_{\mathrm{qSTLS}}(x\to0,l)}{n\beta}=\Theta\frac{x^4}{(2\pi{l}\Theta)^2}\left(\pi\lambda{r}_{\mathrm{s}}\right)\widetilde{u}_{\mathrm{int}}\,.\label{eq:longwavelengthQSTLS}
\end{align}
\textbf{\emph{The short-wavelength limit}}. In the qSTLS auxiliary definite integral, Eq.(\ref{eq:auxqSTLSintegralMatsuNorm}), the integration boundaries not only depend on $x$ but also both become infinite in the short-wavelength limit. This will considerably complicate the calculation. First, it observed that, owing to $x,t\to\infty$, both $(2\pi{l}{\Theta})^2$ contributions are negligible. This leads to
\begin{align*}
\widetilde{A}_{\mathrm{qSTLS}}(x\to\infty,y,w,l)=\int_{x^2-xw}^{x^2+xw}\frac{2dt}{2t+w^2-x^2}\ln{\left|\frac{2xy+t}{2xy-t}\right|}\,.\nonumber
\end{align*}
In view of the integration interval $[x^2-xw,x^2+xw]$, it is straightforward to prove that $\lim_{x\to\infty}(t/x)\to\infty$. Thus, $2xy/t$ constitutes the small argument of interest. The Taylor series $\ln{|(1+x)/(1-x)|}\simeq{2x}+2x^3/3+\cdots$ is employed and only the first order term is retained.
\begin{align*}
\widetilde{A}_{\mathrm{qSTLS}}(x\to\infty,y,w,l)=4xy\int_{x^2-xw}^{x^2+xw}\frac{dt}{t+(w^2-x^2)/2}\frac{1}{t}\,.\nonumber
\end{align*}
The integrand is decomposed into simple fractions. The $\int\,dt/(t+a)=\ln{|t+a|}$ integrals are evaluated. After some algebra, one obtains
\begin{align*}
\widetilde{A}_{\mathrm{qSTLS}}(x\to\infty,y,w,l)=-\frac{8xy}{w^2-x^2}\ln{\left|\frac{x+w}{x-w}\right|}\,.\nonumber
\end{align*}
At this point, $w/x$ constitutes the small argument of interest. The Taylor series $\ln{|(1+x)/(1-x)|}\simeq{2x}+2x^3/3+\cdots$ is again employed and only the first order term is retained. Finally, $w^2-x^2\simeq-x^2$ is employed leading to
\begin{align*}
\widetilde{A}_{\mathrm{qSTLS}}(x\to\infty,y,w,l)=+16wy\frac{1}{x^2}\,.\nonumber
\end{align*}
Overall, the short-wavelength limit of the normalized qSTLS auxiliary Matsubara response becomes
\begin{align}
\frac{\widetilde{I}_{\mathrm{qSTLS}}(x\to\infty,l)}{n\beta}=9\frac{\Theta}{x^2}\int_0^{\infty}w^2\left[S(w)-1\right]dw\int_0^{\infty}dy\frac{y^2}{\exp{\left(\frac{y^2}{\Theta}-\bar{\mu}\right)}+1}\,.\nonumber
\end{align}
We now recall that the microscopic relation between the pair correlation function and the SSF reads as $g(x)=1+(3/2)(1/x)\int_0^{\infty}dyy[S(y)-1]\sin{(xy)}$, in normalized units. This yields $g(0)=1+(3/2)\int_0^{\infty}dyy^2[S(y)-1]$ for the on-top value of the pair correlation function, after the leading term of the Taylor expansion of $\sin{(xy)}$ with $xy\ll1$ is retained, $\sin{(xy)}\simeq{xy}$. Therefore, the short-wavelength limit of the normalized qSTLS auxiliary Matsubara response becomes
\begin{align*}
\frac{\widetilde{I}_{\mathrm{qSTLS}}(x\to\infty,l)}{n\beta}=-6\frac{\Theta}{x^2}\left[1-g(0)\right]\int_0^{\infty}dy\frac{y^2}{\exp{\left(\frac{y^2}{\Theta}-\bar{\mu}\right)}+1}\,.\nonumber
\end{align*}
As per usual, we recall the normalization of the Fermi-Dirac distribution function $\int_0^{\infty}dy\{y^2/[\exp{\left(y^2/\Theta-\bar{\mu}\right)}+1]\}=1/3$. Thus, finally, the long-wavelength limit of the normalized qSTLS auxiliary Matsubara response can be compactly rewritten as
\begin{align}
\frac{\widetilde{I}_{\mathrm{qSTLS}}(x\to\infty,l)}{n\beta}=-2\frac{\Theta}{x^2}\left[1-g(0)\right]\,.\label{eq:shortwavelengthQSTLS}
\end{align}
\textbf{\emph{The high Matsubara frequency limit}}. In the qSTLS auxiliary definite integral, Eq.(\ref{eq:auxqSTLSintegralMatsuNorm}), two different small parameters $\left(2xy+t\right)^2/\left(2{\pi}l\Theta\right)^2$ and $\left(2xy-t\right)^2/\left(2{\pi}l\Theta\right)^2$ can be easily identified. The Taylor series $\ln{|(1\pm{x})|}\simeq\pm{x}-x^2/2+\cdots$ is employed and only the first order term is retained. After some straightforward algebra, one obtains
\begin{align*}
\widetilde{A}_{\mathrm{qSTLS}}(x,y,w,l\to\infty)=\frac{4xy}{\left(2{\pi}l\Theta\right)^2}\int_{x^2-xw}^{x^2+xw}\frac{2tdt}{2t+w^2-x^2}\,.\nonumber
\end{align*}
The term is rearranged and the integration is carried out with the aid of the indefinite integral $\int{dt}/(t+a)=\ln{|t+a|}$. Perfect squares are identified that emerge from both integration boundaries. After some elementary logarithmic algebra, one has
\begin{align*}
\widetilde{A}_{\mathrm{qSTLS}}(x,y,w,l\to\infty)=\frac{4xy}{\left(2{\pi}l\Theta\right)^2}\left[2xw-(w^2-x^2)\ln{\left|\frac{x+w}{x-w}\right|}\right]\,.\nonumber
\end{align*}
Thus, the high Matsubara frequency limit of the normalized qSTLS auxiliary Matsubara response becomes
\begin{align*}
\frac{\widetilde{I}_{\mathrm{qSTLS}}(x,l\to\infty)}{n\beta}=\frac{9}{4}\Theta\frac{x}{\left(2{\pi}l\Theta\right)^2}\int_0^{\infty}w\left[S(w)-1\right]\left[2xw-(w^2-x^2)\ln{\left|\frac{x+w}{x-w}\right|}\right]dw\int_0^{\infty}\frac{y^2dy}{\exp{\left(\frac{y^2}{\Theta}-\bar{\mu}\right)}+1}\,.\nonumber
\end{align*}
As per usual, the normalization of the Fermi-Dirac distribution function is employed $\int_0^{\infty}dy\{y^2/[\exp{\left(y^2/\Theta-\bar{\mu}\right)}+1]\}=1/3$. The terms are rearranged again and the remaining dummy integration variable is switched from $w$ to $y$.
\begin{align*}
\frac{\widetilde{I}_{\mathrm{qSTLS}}(x,l\to\infty)}{n\beta}=-2\Theta\frac{x^2}{\left(2{\pi}l\Theta\right)^2}\left\{-\frac{3}{4}\int_0^{\infty}dyy^2\left[S(y)-1\right]\left[1+\frac{x^2-y^2}{2xy}\ln{\left|\frac{x+y}{x-y}\right|}\right]\right\}\,.\nonumber
\end{align*}
The bracketed term is the normalized form of the static LFC functional of the STLS scheme, see Eq.(\ref{eq:STLSfunctionalLFC}). Thus, finally, the high Matsubara frequency limit of the normalized qSTLS auxiliary Matsubara response can be compactly rewritten as
\begin{align}
\frac{\widetilde{I}_{\mathrm{qSTLS}}(x,l\to\infty)}{n\beta}=-2\Theta\frac{x^2}{\left(2{\pi}l\Theta\right)^2}G_{\mathrm{STLS}}(x)\,.\label{eq:highMatsubaraQSTLS}
\end{align}
\textbf{\emph{The zero Matsubara frequency limit}}.  This is essentially the static limit. The qSTLS auxiliary definite integral, Eq.(\ref{eq:auxqSTLSintegralMatsuNorm}), directly becomes
\begin{align*}
\widetilde{A}_{\mathrm{qSTLS}}(x,y,w,l\to0)=\int_{x^2-xw}^{x^2+xw}\frac{2dt}{2t+w^2-x^2}\ln{\left|\frac{2xy+t}{2xy-t}\right|}\,.\nonumber
\end{align*}
Thus, after some rearrangements in the triple integral, the zero Matsubara frequency limit of the normalized qSTLS auxiliary Matsubara response becomes
\begin{align*}
\frac{\widetilde{I}_{\mathrm{qSTLS}}(x,l\to0)}{n\beta}=\frac{9}{16}\Theta\int_0^{\infty}w\left[S(w)-1\right]dw\int_{x^2-xw}^{x^2+xw}\frac{dt}{2t+w^2-x^2}\int_0^{\infty}\frac{dy}{\exp{\left(\frac{y^2}{\Theta}-\bar{\mu}\right)}+1}\left[2y\ln{\left|\frac{2xy+t}{2xy-t}\right|}\right]\,,\nonumber
\end{align*}
There is a logarithmic singularity at $t=2xy$ which can be removed by integration by parts,
\begin{align*}
\frac{\widetilde{I}_{\mathrm{qSTLS}}(x,l\to0)}{n\beta}=+\frac{9}{16}\Theta\int_0^{\infty}w\left[S(w)-1\right]dw\int_{x^2-xw}^{x^2+xw}\frac{dt}{2t+w^2-x^2}\int_0^{\infty}\frac{dy}{\exp{\left(\frac{y^2}{\Theta}-\bar{\mu}\right)}+1}\frac{\partial}{\partial{y}}\left[\left(y^2-\frac{t^2}{4x^2}\right)\ln{\left|\frac{2xy+t}{2xy-t}\right|}+\frac{y}{x}t\right]\,.\nonumber
\end{align*}
After some algebra, the zero Matsubara frequency limit of the normalized qSTLS auxiliary Matsubara response is finally rewritten as
\begin{align}
\frac{\widetilde{I}_{\mathrm{qSTLS}}(x,l\to0)}{n\beta}=\frac{9}{8}\int_0^{\infty}w\left[S(w)-1\right]dw\int_0^{\infty}dy\frac{y\exp{\left(\frac{y^2}{\Theta}-\bar{\mu}\right)}}{\left[\exp{\left(\frac{y^2}{\Theta}-\bar{\mu}\right)}+1\right]^2}\int_{x^2-xw}^{x^2+xw}\frac{dt}{2t+w^2-x^2}\left[\left(y^2-\frac{t^2}{4x^2}\right)\ln{\left|\frac{2xy+t}{2xy-t}\right|}+\frac{y}{x}t\right]\,,\label{eq:zeroMatsubaraQSTLS}
\end{align}
see also Eq.(\ref{eq:auxqSTLSMatsu0Norm}) of the main text.\\
\textbf{\emph{The combined long-wavelength and zero Matsubara frequency limit}}.  It is understood that the static limit is considered prior to the long-wavelength limit. In the qSTLS auxiliary definite integral, Eq.(\ref{eq:auxqSTLSintegralMatsuNorm}), the change of variables $t=x^2-xwu$ is employed, which transforms the integration interval to $[-1,+1]$ and also leads to the differential $dt=-xwdu$.
\begin{align*}
\widetilde{A}_{\mathrm{qSTLS}}(x\to0,y,w,l\to0)&=\frac{2xw}{w^2}\int_{-1}^{+1}\frac{du}{1-(2xwu-x^2)/w^2}\ln{\left|\frac{x-wu+2y}{x-wu-2y}\right|}\,.\nonumber
\end{align*}
The small argument for the pre-factor is identified to be $(2xwu-x^2)/w^2$, the small argument for the first logarithm is identified to be $x/(wu-2y)$ and the small argument for the second logarithm is identified to be $x/(wu+2y)$. In the Taylor expansions of the three terms only the first order term with respect to $x$ is retained. After some cumbersome algebra, the above lead to
\begin{align*}
\widetilde{A}_{\mathrm{qSTLS}}(x\to0,y,w,l\to0)&=\frac{4x^2}{w^2}\int_{-1}^{+1}duu\ln{\left|\frac{u-2y/w}{u+2y/w}\right|}+\frac{2x^2}{w^2}\int_{-1}^{+1}du\left(\frac{1}{u+2y/w}-\frac{1}{u-2y/w}\right)\,.\nonumber
\end{align*}
The definite integrations are carried out based on the integrals $\int\,du/(u+a)=\ln{|u+a|}$ and $\int\,udu\ln{|(u+a)/(u-a)|}=[(u^2-a^2)/2]\ln{|(u+a)/(u-a)|}+au$. The two contributions are combined to yield
\begin{align*}
\widetilde{A}_{\mathrm{qSTLS}}(x\to0,y,w,l\to0)&=-\frac{16x^2}{wy}\left[\frac{y^2}{w^2}-\frac{y^3}{w^3}\ln{\left|\frac{w+2y}{w-2y}\right|}\right]\,.\nonumber
\end{align*}
Overall, the long-wavelength limit of the static normalized qSTLS auxiliary Matsubara response becomes
\begin{align}
\frac{\widetilde{I}_{\mathrm{qSTLS}}(x\to0,l\to0)}{n\beta}=-9\Theta{x}^2\int_0^{\infty}\left[S(w)-1\right]dw\int_0^{\infty}\frac{dy}{\exp{\left(\frac{y^2}{\Theta}-\bar{\mu}\right)}+1}\left[\frac{y^2}{w^2}-\frac{y^3}{w^3}\ln{\left|\frac{w+2y}{w-2y}\right|}\right]\,.\label{eq:longwavelengthzeroMatsubaraQSTLS}
\end{align}

\bibliography{main_bib}

\end{document}